\documentstyle[12pt,epsfig]{article}

\def\journal#1#2#3#4{{\it #1} {\bf #2} (#3) #4}

\def\prl{Phys. Rev. Lett.}
\def\pl{Phys. Lett.}
\def\np{Nucl. Phys.}

\def\pr{Phys. Rev.}

\def\jhep{JHEP}

\def\ctp{Commun. Theor. Phys.}
\def\rmp{Rev. Mod. Phys.}
\def\ml{{\hat{m_{\ell}}}}
\def\ddp{{D^\prime}}
\def\mc{{\hat{m_c}}}
\def\c{C}
\def\cs{{\c_7}}
\def\cn{{\c_9}}
\def\ct{{\c_{10}}}
\def\cne{\cn^{\rm eff}}
\def\cse{\cs^{\rm eff}}
\def\m{{\cal M}}
\def\gl{\Gamma}
\def\g{\gamma}
\def\l{\ell}
\def\ks{{K^\ast}}
\def\lb{\bar{\l}}

\def\bkll{B \rightarrow K \, \l^+ \, \l^-}
\def\bksll{B \rightarrow \ks \, \l^+ \, \l^-}

\def\bkm{B \rightarrow K \mu^+ \mu^-}
\def\bkt{B \rightarrow K \tau^+ \tau^-}
\def\bksm{B \rightarrow K^* \mu^+ \mu^-}
\def\bkst{B \rightarrow K^* \tau^+ \tau^-}

\def\ph{{p_h}}

\def\he{{\cal H}_{\rm eff}}

\def\d{{\rm d}}
\def\D{{\cal D}}

\def\t{{\rm T}}

\def\mh{\hat{m}}
\def\mbh{\mh_b}
\def\msh{\mh_s}
\def\mph{\mh_K}
\def\mvh{\mh_{K^*}}

\def\mlh{\mh_\l}
\def\qh{\hat{q}}
\def\pvh{\hat{p}_{K^*}}

\def\pbh{\hat{p}_B}
\def\ph{\hat{p}}
\def\sh{\hat{s}}

\def\t{{\cal T}}
\def\a{{\cal A}}
\def\s{{\cal S}}
\def\cqb{C_{Q1}}
\def\cqc{C_{Q2}}
\def\ep{{\epsilon^\ast}}
\def\ap{{A^\prime}}

\def\cp{{C^\prime}}
\def\rp{{D^\prime}}
\def\uh{{\hat{u}}}
\def\la{{\lambda}}

\def\be{\begin{equation}}
\def\ee{\end{equation}}
\def\ba{\begin{eqnarray}}
\def\ea{\end{eqnarray}}
\def\nnb{\nonumber}
\begin{document}
\renewcommand{\thefootnote}{\fnsymbol{footnote}}

\begin{titlepage}

\begin{flushright}
\begin{tabular}{l}
TUHEP-TH-00116\\
hep-ph/0004262
\end{tabular}
\end{flushright}
\vskip0.5cm
\begin{center}
{\LARGE\bf
Exclusive Semileptonic Rare
Decays $B \to (K,K^*) \ell^+ \ell^-$ in Supersymmetric Theories}

\vspace*{0.5cm}

             {\bf Qi-Shu YAN}\footnote{
        E-mail : yanqs@itp.ac.cn; qsyan@mail.tsinghua.edu.cn}\\

	\vspace{0.5cm}
             Physics Department of Tsinghua University,P.R.China \\

       \vspace{0.5cm}

       {\bf Chao-Shang HUANG}\footnote{
        E-mail : csh@itp.ac.cn},
	{\bf LIAO Wei}\footnote{
	E-mail: liaow@itp.ac.cn},
       {\bf Shou-Hua ZHU}\footnote{
	E-mail: huald@itp.ac.cn}
        \vspace{0.5cm}

        Institute of Theoretical Physics,\\
        the Chinese Academy Science, P.R.China

        \bigskip



\bigskip

  {\bf Abstract\\[10pt]}
\end{center}
The invariant mass spectrum, forward-backward asymmetry, and
lepton  polarizations of the exclusive processes $B\rightarrow
K(K^*)\ell^+ \ell^-, \ell=\mu,~\tau$ are analyzed under supersymmetric
context. Special attention is paid to the effects of neutral Higgs
bosons (NHBs). Our analysis shows that the branching ratio of the process $\bkm$
can be quite largely modified by the effects of neutral Higgs bosons and the forward-backward asymmetry
would not vanish. For the process $\bksm$, the lepton transverse polarization is quite sensitive
to the effects of NHBs, while the invariant mass spectrum, forward-backward
asymmetry, and lepton longitudinal polarization are not. For both $\bkt$ and $\bkst$, the effects of NHBs are
quite significant. The partial decay widths of these processes are also analyzed,
and our analysis manifest that even taking into account the theoretical uncertainties in calculating
weak form factors, the effects of NHBs could make SUSY shown up.

\vfill

\bigskip

\centerline{{\sc Pacs} numbers: 13.20.He, 13.25.Hw, 12.60.Jv}
\end{titlepage}
\renewcommand{\thefootnote}{\arabic{footnote}}
\setcounter{footnote}{0}

\section{\bf Introduction}

The inclusive rare processes $b\rightarrow X_s \ell^+ \ell^-, \ell=e,~\mu,~\tau$
have been intensively studied in
literatures\cite{gsw,ex1,dhh,hly,fky,fkmy,kkl,goto96,lmss99,ht96,gln,ks}.
As one of flavor changing neutral current processes, it
is sensitive to fine structure of the standard model and to the possible
new physics as well, and is expected to shed light on the existence of new
physics before the possible new particles are produced at colliders.

It is well known that invariant mass spectrum, forward-backward asymmetries(FBAs), and lepton
polarizations are important observables to probe new physics, while the first two
observables are mostly analyzed. About lepton polarizations, it is known that due to the smallness
of the mass of it, therefore electron polarizations are very difficult to be measured
experimentally. So only the lepton polarizations of muon and tau are
considered in literatures~\cite{ht96,ks,ch,hz}. The longitudinal polarization of tau in $B \rightarrow X_s \tau^+
 \tau^-$ has been calculated in standard model (SM) and several new physics scenarios~\cite{ht96}.  For
$B\rightarrow X_s l^+ l^-$ ($l =\mu, \tau$), the polarizations of lepton in SM are
analyzed in \cite{ks} and it is pointed out that for the $\mu$ channel, the only
significant component is the longitudinal polarization($P_L$), while all three components are sizable in the
$\tau$ channel.The analysis has been extended to supersymmetric models (SUSY) and a CP softly broken two Higgs doublet
model in refs. ~\cite{ch} and \cite{hz} respectively. The reference \cite{fky} also gives a general model-independent
analysis of the lepton polarization asymmetries in the process $B \rightarrow
X_s \tau^+ \tau^-$ and it is found that the contribution from
$C_{LRLR}+C_{LRRL}$ is much larger than other scalar-type interactions.

Compared with the inclusive processes $B\rightarrow X_s \ell^+ \ell^-,
\ell=e,~\mu,~\tau$, the theoretical study of the exclusive processes
$B\rightarrow K(K^*) \ell^+ \ell^-$ is relatively hard. For inclusive semileptonic decays of B, the decay
rates can be calculated in heavy quark effective theory (HQET) \cite{hqet}. However,
for exclusive semileptonic decays of B, to make theoretical predictions, additional
knowledge of decay form factors is needed, which is related with the
calculation of hadronic transition matrix elements. Hadronic transition matrix
elements depend on the non-perturbative properties of QCD, and can only
be reliably calculated by using a nonperturbative method. The form factors for B decay into $K^{(*)}$ have been computed with
different methods such as quark models~\cite{jw}, SVZ  QCD sum rules~\cite{cfss}, light cone sum rules
(LCSRs)~ \cite{bbk,chzh,brau,kr,ball}. Compared to the lattice approach which
mainly deal with the form factors at small recoil, the QCD sum rules on the
light-cone can complementarily provide information of the form factors
at smaller values of $\sh$. And they are consistent with perturbative QCD and
the heavy quark limit. In this work, we will use the weak decay form factors
calculated by using the technique of the light cone QCD sum rules and given in
\cite{abhh}.

A upper limit on the branching ratio of $B^0\rightarrow K^{0*} \mu^+\mu^-$  has been recently given by CLEO \cite{cl}:
\begin{equation}
BR ( B^o \rightarrow K^{0*} \mu^+\mu^-) < 4.0 \times 10^{-6} ,
\end{equation}
and  they will be
precisely measured at B factories, these exclusive processes are quite worthy
of intensive study and have attracted many attentions \cite{abhh,krp,dt,ex,dl,liu,akk,ju,mns,giw,as,asok,gk}.
In reference \cite{abhh}, by
using improved theoretical calculations of the decay form factors in the light
cone QCD sum rule approach, dilepton invariant mass spectra and the
FBAs of these exclusive decays are analyzed in the
standard model and a number of popular variants of the supersymmetric models.
However, as the author claimed, the effects of neutral Higgs exchanges are
neglected. For exclusive processes, as pointed out in
\cite{gk}, the polarization asymmetries of $\mu$ and $\tau$ for $B \rightarrow
K^* \mu^+ \mu^-$ and $ B \rightarrow K^* \tau^+ \tau^-$ are also accessible
at the B-factories under construction.  In reference \cite{as}, the lepton polarizations and CP violating effects in
 $B \rightarrow K^* \tau^+ \tau^-$ are analyzed in SM and two Higgs doublet models.

As pointed in refs.~\cite{dhh,hly}, in two-Higgs-doublet models and SUSY models, neutral Higgs boson could contribute
largely to the inclusive processes $B\rightarrow X_s \ell^+ \ell^-,
\ell=~\mu,~\tau$ and greatly modify the branching ratio and FBA in the large tan$\beta$
case. The effects of
neutral Higgs in the 2HDM to polarizations of $\tau$ in $B\rightarrow K \tau^+
\tau^-$ are analyzed in \cite{asok}, and it was found that polarizations of the charged
final lepton are very sensitive to the tan$\beta$.

In this paper, we will investigate the exclusive decay $B\rightarrow K(K^*) \ell^+ \ell^-, \ell=\mu,~\tau$
in SUSY models. We shall evaluate branching ratios and FBAs with emphasis on the
effects of neutral Higgs  and analyze lepton polarizations in MSSM.
According to the analysis of \cite{liu}, different sources of the vector current
could manifest themselves in different regions of phase space, for the very
low $\sh$ the photonic penguin dominates, while the Z penguin and W box
becomes important towards high $\sh$. In order to search the regions of $\sh$ where
neutral Higgs bosons could contribute large, we analyze the partial
decay widths of these two processes. Beside that they
are accessible to B factories, our motivation also bases on the fact that to
the inclusive processes $B\rightarrow X_s \ell^+ \ell^-, \ell=\mu, ~\tau$,
neutral Higgs could make quite a large contributions at certain large tan$\beta$ regions
of parameter space in SUSY models, since part of supersymmetric contributions
is proportional to tan$^3\beta$~ \cite{hly}. Such regions considerably exist in SUGRA
and M-theory inspired models~\cite{hllyz}.
We also analyze the effects of neutral
Higgs to the position of the zero value of the FBA.
Our results show that the branching ratio of the process $\bkm$
can be quite largely modified by the effects of neutral Higgs bosons and the FBA
would not vanish. Because the FBA for $\bkll (\ell=\mu,~\tau)$ vanishes
if the contributions of NHBs vanish. The contributions of NHBs can
be large enough to be observed only in SUSY and/or 2HDM with large tan$\beta$, and a non-zero FBA for $\bkll$
would signal the existance of new physics.
For the process $\bksm$,  the lepton transverse polarization is quite sensitive
to the effects of NHBs, while the invariant mass spectrum, FBA
, and lepton longitudinal polarization are not. For both $\bkt$ and $\bkst$, the effects of NHBs are
quite significant. Our analysis manifest that even taking into account the theoretical uncertainties in calculating
weak form factors, the effects of NHBs could show SUSY up. In brief, our analysis manifest
that effects of NHBs is quite remarkable in some regions of parameter space
of SUSY, even for the process $\bkm$.

The paper is organized as follows. In the section 2, the effective
Hamiltonian is presented and the form factors given by using light cone  sum rule method are briefly
discussed. Basic formula of observables are introduced in section 3. Section 4 is devoted to the numerical
analysis. In section 5 we make discussions and conclusions.

\section{\bf Effective Hamiltonian and Form Factors}
\setcounter{equation}{0}

By integrating out the degrees of heavy freedom from the
full theory, MSSM, at electroweak(EW) scale, we can get the effective Hamiltonian
describing the rare semileptonic decay  $b \to s  \ell^+ \ell^-$:
\begin{equation}
\he  =  -\frac{4G_{F}}{\sqrt{2}} V_{tb} V_{ts}^{*} (\sum_{i=1}^{10}
    C_{i}(\mu) O_{i}(\mu) + \sum_{i=1}^{10} C_{Q_i}(\mu) Q_i(\mu))\; ,
        \label{eq:he}
\end{equation}

where the first ten operators and Wilson coefficients at EW scale can be found in
\cite{goto96,bbmr}\footnote{In our previous papers, e.g., \cite{dhh,hly}, we follow the convention of
ref. \cite{gsw} for the indices of operators as well as Wilson coefficients. In this paper.
we use more popular conventions (see, e.g., \cite{efh}). That is, $O_8\rightarrow O_9$
and $O_9\rightarrow O_{10}$.}, and last ten operators and Wilson coefficients which represent the contributions of
neutral Higgs can be found in \cite{hly}.

With the renormalization group equations to resum the QCD corrections, Wilson coefficients at
energy scale $\mu=m_b$ are evaluated. Theoretical uncertainties
related to renormalization-scale can be substantially reduced when the
next-leading-logarithm corrections are included \cite{efh}.

The above Hamiltonian leads to the following free quark decay amplitude:
\begin{eqnarray}         \m(b\to s\ell^+\ell^-) & = & -\frac{G_F
\alpha}{\sqrt{2}  \pi} \,                  V_{t s}^\ast V_{tb} \, \left\{
             \cne  \left[ \bar{s}  \g_\mu  L  b \right] \,
       \left[ \lb  \g^\mu  \l \right]                 + \ct  \left[ \bar{s}
\g_\mu  L  b \right] \,                           \left[ \lb  \g^\mu  \g_5  \l
\right]                 \right. \nonumber \\         & & \left.
                - 2 \mbh  \cse  \left[ \bar{s}  i  \sigma_{\mu \nu}
                        \frac{\qh^{\nu}}{\sh}  R  b \right]
                        \left[ \lb  \g^\mu  \l \right]
                +\cqb \left[ \bar{s}  R  b \right] \,
                         \left[ \lb \l \right]
                +\cqc \left[ \bar{s}  R  b \right] \,
                         \left[ \lb \g_5 \l \right]\right\} \; .
        \label{eq:m}
\end{eqnarray}

where $C^{eff}_9$ is defined as \cite{agm,bm}
\begin{equation}
\cne (\mu, \hat{s}) = C_9(\mu) + Y(\mu, \sh) + \frac{3 \pi}{\alpha^2} C (\mu)
         \sum_{V_i = \psi(1s),..., \psi(6s)} \kappa_i
      \frac{\Gamma(V_i \rightarrow \ell^+ \ell^-)\, m_{V_i}}{
      {m_{V_i}}^2 - \sh \, {m_B}^2 - i m_{V_i} \Gamma_{V_i}}
\label{eqn:cni}
\end{equation}
where $\hat{s}=s/m_b^2$,s=$q^2$, $C(\mu)=\left(3 \, C_1 + C_2 + 3 \, C_3
                + C_4 + 3 \, C_5 + C_6 \right)$, and

\ba
        {Y}(\mu, \sh) & = & g(\mc,\sh) C(\mu) \nonumber \\
         &&  -\frac{1}{2} g(1,\sh)
                \left( 4 \, C_3 + 4 \, C_4 + 3 \,
                C_5 + C_6 \right)
         - \frac{1}{2} g(0,\sh) \left( C_3 +
                3 \, C_4 \right) \nonumber \\
        &&-     \frac{2}{9} \left( 3 \, C_3 + C_4 +
                3 \, C_5 + C_6 \right) \; .
\label{eq:yert}
\ea
where the function $g(\mc, \sh)$ comes from one loop contributions of four-quark operators and is defined by
\begin{equation}
g(z,\hat{s})=-\frac{4}{9} lnz^2+ \frac{8}{27}+ \frac{16}{9}
\frac{z^2}{\hat{s}}-\left\{ \begin{array}{ll}
 \frac{2}{9} \sqrt{1-\frac{4 z^2}{\hat{s}}} (2+\frac{4 z^2}{\hat{s}})
 \bigg[ ln ( \frac{1+\sqrt{1-4 z^2/ \hat{s}}} {1-\sqrt{1-4 z^2/ \hat{s}}})
 -i \pi \bigg], & \textrm{$4 z^2 < \hat{s}$} \\
 \frac{4}{9} \sqrt{\frac{4 z^2}{\hat{s}}-1} (2+\frac{4 z^2}{\hat{s}})
arctan \bigg( \frac{1}{\sqrt{4 z^2/ \hat{s} -1}} \bigg),
 & \textrm{$4 z^2 > \hat{s}$}
\end{array} \right.
\end{equation}
The last terms in
(\ref{eqn:cni}) are nonperturbative effects from $(\bar{c} c)$ resonance contributions,
while the phenomenological factors $\kappa_i$ can be fixed from the processes \cite{abhh}
$B\to K^{(\ast)} V_i \to K^{(\ast)} \ell^+ \ell^-$ and as given in the Table. \ref{tab:kappa}.
\begin{table}[b]
        \begin{center}
        \begin{tabular}{|c|c|c|}
 \hline
    \multicolumn{1}{|c|}{$\kappa$}
      & \multicolumn{1}{|c|}{$J/\Psi$}
      & \multicolumn{1}{|c|}{$\Psi^\prime$} \\
        \hline
  $K$           &2.70                   & 3.51 \\
  $K^\ast$      &1.65                   & 2.36 \\
        \hline
        \end{tabular}
        \end{center}
\caption{\it Fudge factors in $B\to K^{(\ast)} J/\Psi, \Psi^\prime \to
K^{(\ast)} \ell^+ \ell^- $ decays calculated using the LCSR form factors.}
\label{tab:kappa}
\end{table}

Exclusive decays $B\to (K,K^*) \ell^+ \ell^-$ are described in terms
of matrix elements of the quark operators in Eq. (\ref{eq:m}) over meson
states, which can be parametrized in terms of form factors.

For the process $B\to K \ell^+ \ell^-$, the non-vanishing matrix
elements are ($q=p_B-p$)
\begin{equation}\label{eq:ff1}
\langle K(p) | \bar s \gamma_\mu b | B(p_B)\rangle  =  f_+(s) \left\{
(p_B+p)_\mu - \frac{m_B^2-m_K^2}{s} \, q_\mu \right\} +
\frac{m_B^2-m_K^2}{s} \, f_0(s)\, q_\mu,
\end{equation}
and
\begin{eqnarray}
\langle K(p) | \bar s \sigma_{\mu\nu} q^\nu (1+\gamma_5) b | B(p_B)\rangle
& = &  \langle K(p) | \bar s \sigma_{\mu\nu} q^\nu b |
B(p_B)\rangle\nonumber\\
& = & i\left\{ (p_B+p)_\mu s - q_\mu (m_B^2-m_K^2)\right\} \,
  \frac{f_T(s)}{m_B+m_K}.\label{eq:ff2}
\end{eqnarray}
While for $B\to K^* \ell^+ \ell^-$, related transition matrix elements are
\begin{eqnarray}
\langle K^*(p) | (V-A)_\mu | B(p_B)\rangle & = & -i \epsilon^*_\mu
(m_B+m_{K^*}) A_1(s) + i (p_B+p)_\mu (\epsilon^* p_B)\,
\frac{A_2(s)}{m_B+m_{K^*}}\nonumber\\
\lefteqn{+ i q_\mu (\epsilon^* p_B) \,\frac{2m_{K^*}}{s}\,
\left(A_3(s)-A_0(s)\right) +
\epsilon_{\mu\nu\rho\sigma}\epsilon^{*\nu} p_B^\rho p^\sigma\,
\frac{2V(s)}{m_B+m_{K^*}}\,.}\hspace*{2cm}\label{eq:ff3}
\end{eqnarray}
and
\begin{eqnarray}
\langle {K^*} | \bar s \sigma_{\mu\nu} q^\nu (1+\gamma_5) b |
B(p_B)\rangle & = & i\epsilon_{\mu\nu\rho\sigma} \epsilon^{*\nu}
p_B^\rho p^\sigma \, 2 T_1(s)\nonumber\\
& & {} + T_2(s) \left\{ \epsilon^*_\mu
  (m_B^2-m_{{K^*}}^2) - (\epsilon^* p_B) \,(p_B+p)_\mu \right\}\nonumber\\
& & {} + T_3(s)
(\epsilon^* p_B) \left\{ q_\mu - \frac{s}{m_B^2-m_{{K^*}}^2}\, (p_B+p)_\mu
\right\}\label{eq:T}
\end{eqnarray}

Where $\epsilon_\mu$ is polarization vector of the vector meson $K^*$. By means of the equation of motion,
one obtains  several relations between form factors
\begin{eqnarray}  A_3(s) & = &
\frac{m_B+m_{K^*}}{2m_{K^*}}\, A_1(s) - \frac{m_B-m_{K^*}}{2m_{K^*}}\,
A_2(s),\nonumber\\ A_0(0) & = & A_3(0), \nonumber\\
\langle K^* |\partial_\mu A^\mu | B\rangle & = & 2 m_{K^*}
(\epsilon^* p_B) A_0(s),\nonumber \\
T_1(0) &= & T_2(0).
\label{eq:exq}
\end{eqnarray}
All signs are defined in such a way as to render the form factors real and positive. The physical
range in $\sh$ extends from $\sh_{\rm min} = 4 \mlh^2$ to $\sh_{\rm max} = (1-{\hat m}_{K,K^*})^2$.

The calculation of the form factors given above is a real task, and one has to
rely on certain approximate methods. We use the results calculated by using technique
of LCSRs and given in \cite{abhh}. And form factors can be parametrized as
\begin{equation}\label{eq:para}
F(\hat{s}) = F(0) \exp ( c_1 \hat{s} + c_2 \hat{s}^2 + c_3 \hat{s}^3).
\end{equation}
The parameterization formula works within 1\% accuracy for $s<15\,$GeV$^2$ and
can avoid the spurious singularities at $s=m^2_{B}$. Related parameters is
given in the Table. 4 of \cite{abhh}

\section{\bf Formula of Observables}
\setcounter{equation}{0}

In this section we provide formula of experimental observables, which include
dilepton invariant mass spectrum, FBA, and lepton polarizations.

>From eqs. (2.2 - 2.8), it is straightforward to obtain the matrix element of $B\rightarrow K (K^{*}) l^+l^-$ as follows.
\begin{equation}
        \m  =  -\frac{G_F  \alpha}{2 \sqrt{2} \pi} \,
                V_{ts}^\ast  V_{tb}  m_B \, \left[
                  \t_\mu^1 \, \left( \lb \, \g^\mu \, \l \right)
                + \t_\mu^2 \,
                  \left( \lb \, \g^\mu \, \g_5 \, \l \right)
                +\s  \left( \lb \l \right ) \right] \; ,
        \label{eq:med}
\end{equation}
where for $B\to K\ell^+\ell^-$,
\begin{eqnarray}
  \t_\mu^1 & = & \ap(\sh) \, \ph_\mu,
   \label{eq:t1bpll}\\
  \t_\mu^2 & = & \cp(\sh) \, \ph_\mu + \rp(\sh) \, \qh_\mu \; ,
     \label{eq:t2bpll}\\
  \s & = & \s_1(\sh)
     \label{eq:sbpll}\,
\end{eqnarray}
and for $B\to K^*\ell^+\ell^-$,
\begin{eqnarray}
  \t_\mu^1 & = &
    A(\sh) \, \epsilon_{\mu\rho\alpha\beta} \ep^\rho \pbh^\alpha
    \pvh^\beta
    - i B(\sh) \, \ep_\mu
    + i C(\sh) \, (\ep \cdot \hat{p}_B) \ph_\mu  \; ,
    \label{eq:t1bvll}\\
  \t_\mu^2 & = &
  E(\sh) \, \epsilon_{\mu\rho\alpha\beta} \ep^\rho \pbh^\alpha \pvh^\beta
  - i F(\sh) \, \ep_\mu
    + i G(\sh) \, (\ep \cdot \hat{p}_B) \ph_\mu
    + i H(\sh) \, (\ep \cdot \hat{p}_B) \qh_\mu \; ,
    \label{eq:t2bvll}\\
  \s & = &i 2 \mvh (\ep \cdot \hat{p}_B) \s_2(\sh)
\end{eqnarray}
with $p \equiv p_B + p_{K,K^*}$. Note that, using the equation of
motion for lepton fields, the terms in $\hat{q}_\mu$
in ${\cal T}^1_\mu$ vanish.

The auxiliary functions above are
defined as
\begin{eqnarray}
  \ap(\sh) & = & \cne(\sh) \, f_+(\sh)
         + \frac{2 \mbh}{1 + \mph} \cse f_T(\sh) \;, \label{eq:axb} \\
  \cp(\sh) & = & \ct \, f_+(\sh) \; , \\
  \rp(\sh) & = & \ct \, f_-(\sh) -\frac{1-\mph^2}{2 \mlh (\mbh-\msh)} \cqc
f_0(\sh) \; , \\
  \s_1(\sh)  &=& \frac{1-\mph^2}{(\mbh-\msh)} \cqb f_0(\sh)
\;,\\   A(\sh) & = & \frac{2}{1 + \mvh} \cne(\sh) V(\sh)
         + \frac{4 \mbh}{\sh} \cse T_1(\sh) \; , \\
  B(\sh) & = & (1 + \mvh) \left[ \cne(\sh) A_1(\sh)
         + \frac{2 \mbh}{\sh} (1 - \mvh) \cse T_2(\sh) \right] \; , \\
  C(\sh) & = & \frac{1}{1 - \mvh^2} \left[
         (1 - \mvh) \cne(\sh) A_2(\sh)
         + 2 \mbh \cse \left(
           T_3(\sh) + \frac{1 - \mvh^2}{\sh} T_2(\sh) \right) \right] \; , \\
  E(\sh) & = & \frac{2}{1 + \mvh} \ct V(\sh) \; , \label{eq:dt}\\
  F(\sh) & = & (1 + \mvh) \ct A_1(\sh) \; , \\
  G(\sh) & = & \frac{1}{1 + \mvh} \ct A_2(\sh) \; , \\
  H(\sh) & = & \frac{\ct}{\sh} \left[
       (1 + \mvh) A_1(\sh) - (1 - \mvh) A_2(\sh) - 2 \mvh A_0(\sh) \right]
\nonumber \\
&&+\frac{\mvh}{\mlh (\mbh+\msh)} A_0(\sh) \cqc \; , \\
\s_2(\sh)&=&-\frac{1}{(\mbh+\msh)} A_0(\sh) \cqb \;. \label{eq:axd}
\end{eqnarray}
where
\begin{eqnarray}
f_0(\sh)&=&\frac{1}{1-\mph^2} [\sh f_-(\sh)+(1-\mph^2) f_+(\sh)]
\end{eqnarray}
and to get the auxiliary functions given above, we have used equations of motion
\begin{eqnarray}
q^{\mu} (\bar \psi_1 \g_\mu \psi_2)&=&(m_1-m_2) \bar \psi_1 \psi_2, \\
q^{\mu} (\bar \psi_1 \g_\mu \g_5 \psi_2)&=&-(m_1+m_2) \bar \psi_1 \g_5 \psi_2.
\end{eqnarray}
The contributions of NHBs have been incorporated in the terms of $\s_1(\sh)$, $D^{\prime}(\sh)$,
$H(\sh)$ and $\s_2(\sh)$. It is remarkable that the contributions of NHBs in
$D^{\prime}(\sh)$ and $H(\sh)$ are proportional to the inverse mass of the lepton, and
for the case $l=\mu$, the effects of NHBs can be manifested through these terms.

A phenomenological effective Hamiltonian is recently given in \cite{akk}.
If ignoring tensor type interactions in the phenomenological Hamiltonian (it is shown that physical observables
are not sensitive to the presence of tensor type interactions~\cite{fkmy}), it is easy to verify that  the matrix
element of $B\rightarrow K^{(*)}\ell^+\ell^-$ can always be
expressed as the form of the equation (\ref{eq:med}) with the auxiliary functions  defined as
\ba
\ap(\sh) & = & w_{c_1} \, f_+(\sh)
         - \frac{w_{c_9}+w_{c_{10}}}{1+\mph} f_T(\sh) \; , \label{eq:auxb} \\
  \cp(\sh) & = & w_{c_2} \, f_+(\sh)\; , \\
  \rp(\sh) & = & w_{c_2} \, f_-(\sh) -\frac{1-\mph^2}{2 \mlh (\mbh-\msh)} w_{c_6}
f_0(\sh) \; , \\
  \s_1(\sh)  &=& \frac{1-\mph^2}{(\mbh-\msh)} w_{c_5} f_0(\sh)
\;,\\   A(\sh) & = & \frac{2}{1 + \mvh} w_{c_1} V(\sh)
         - \frac{2)}{\sh} (w_{c_9}+w_{c_{10}}) T_1(\sh) \; , \\
  B(\sh) & = & -(1 + \mvh) \left[ w_{c_3} A_1(\sh)
         + \frac{1}{\sh} (1 - \mvh) (w_{c_9}+w_{c_{10}})  T_2(\sh) \right] \; , \\
  C(\sh) & = & -\frac{1}{1 - \mvh^2} \left[
         (1 - \mvh) w_{c_3} (\sh) A_2(\sh) \right .\nnb \\
&&\left .+ (w_{c_9}-w_{c_{10}}) \left(
           (1+\mvh) T_3(\sh) + \frac{1 - \mvh^2}{\sh} T_2(\sh) \right) \right] \; , \\
  E(\sh) & = & \frac{2}{1 + \mvh} w_{c_2} V(\sh) \; , \\
  F(\sh) & = & -(1 + \mvh) w_{c_4} A_1(\sh) \; , \\
  G(\sh) & = & -\frac{1}{1 + \mvh} w_{c_4} A_2(\sh) \; , \\
  H(\sh) & = & -\frac{2 \mvh}{\sh}\,w_{c_4}\, \left( A_3(\sh)-A_0(\sh) \right)+\frac{\mvh}{\mlh (\mbh+\msh)} w_{c_8} A_0(\sh)\; , \\
 \s_2(\sh)&=&-\frac{1}{(\mbh+\msh)} w_{c_7} A_0(\sh)  \;,
\label{eq:auxd}
\ea
where
\ba
w_{c_1} &=& \frac{1}{4} (C_{LL}+C_{LR}+C_{RL}+C_{RR}) \;, \\
w_{c_2} &=& \frac{1}{4} (-C_{LL}+C_{LR}-C_{RL}+C_{RR})\;, \\
w_{c_3} &=& \frac{1}{4} (-C_{LL}-C_{LR}+C_{RL}+C_{RR})\;, \\
w_{c_4} &=& \frac{1}{4} (C_{LL}-C_{LR}-C_{RL}+C_{RR}) \;, \\
w_{c_5} &=& \frac{1}{4} (C_{LRLR}+C_{RLLR}+C_{LRRL}+C_{RLRL})  \;, \\
w_{c_6} &=& \frac{1}{4} (C_{LRLR}+C_{RLLR}-C_{LRRL}-C_{RLRL})\;, \\
w_{c_7} &=& \frac{1}{4} (C_{LRLR}-C_{RLLR}+C_{LRRL}-C_{RLRL}) \;, \\
w_{c_8} &=& \frac{1}{4} (C_{LRLR}-C_{RLLR}-C_{LRRL}+C_{RLRL}) \;, \\
w_{c_9} &=& m_b C_{BR}    \;, \\
w_{c_{10}} &=& m_s C_{SL}   \;.
\ea
In the above equations $C_{LL}, C_{LR}$ etc are defined in ref. ~\cite{fkmy}.
Therefore our formula given below can also be used to make  model independent phenomenological analysis,
if using Eqs. ((\ref{eq:auxb})-(\ref{eq:auxd})) instead of Eqs. ((\ref{eq:axb})-(\ref{eq:axd})).

Keeping the lepton mass, we find the double differential
decay widths $\gl^K$ and $\gl^{K^*}$ for the decays $B\to
K\ell^+\ell^-$ and $B\to K^*\ell^+\ell^-$, respectively, as
\begin{eqnarray}
  \frac{\d^2 \gl^K}{\d\sh \d\uh} & = &
  \frac{G_F^2  \alpha^2  m_B^5}{2^{11}  \pi^5}
      \left| V_{ts}^\ast  V_{tb} \right|^2  \nonumber \\
& & \times\left\{
(|\ap|^2 +|\cp|^2)
(\la -\uh^2 ) +|\s_1|^2 (\sh-4 \mlh^2)+Re(\s_1 \ap^{*}) 4 \mlh  \uh
\right. \nonumber \\
&& + \left. |\cp|^2 4 \ml^2 (2+2 \mph^2-\sh)
+Re( \cp \ddp^{*}) 8 \ml^2 (1-\mph^2)
+|\ddp|^2 4 \ml^2 \sh \right\} \; ,    \label{eq:ddwbpll} \\
  \frac{\d^2 \gl^{K^*}}{\d\sh \d\uh} & = &
  \frac{G_F^2 \, \alpha^2 \, m_B^5}{2^{11} \pi^5}
      \left| V_{ts}^\ast \, V_{tb} \right|^2
        \nonumber \\
  & &  \times\Bigg\{
  \frac{|A|^2}{4}
   \left(\sh (\la + \uh^2) + 4 \mlh^2 \la  \right)
  + \frac{|E|^2}{4} \left(\sh (\la + \uh^2) - 4 \mlh^2 \la  \right)
  +|\s_2|^2 (\sh-4\mlh^2) \la
        \Bigg.
        \nonumber \\
  & & \Bigg.
  + \frac{1}{4 \mvh^2} \left[
  |B|^2 \left( \la - \uh^2 + 8 \mvh^2 (\sh +2 \mlh^2 ) \right)
  + |F|^2 \left( \la - \uh^2 + 8 \mvh^2 (\sh -4 \mlh^2) \right) \right]
        \Bigg.
        \nonumber \\
  & & \Bigg.
  - 2 \sh \uh \left[ {\rm Re}(BE^\ast) + {\rm Re}(AF^\ast)
\right]+\frac{2 \mlh \uh}{\mvh}[{\rm Re}   (\s_2 B^*) (\sh+\mvh^2-1) + {\rm
Re}(\s_2 C^*) \la ]        \Bigg.
        \nonumber \\
  & & \Bigg.
  + \frac{\la}{4 \mvh^2} \left[ |C|^2 (\la - \uh^2)
    + |G|^2 \left( \la - \uh^2 + 4 \mlh^2 (2 + 2 \mvh^2 - \sh) \right) \right]
        \Bigg.
        \nonumber \\
  & & \Bigg.
  - \frac{1}{2 \mvh^2} \left [
  {\rm Re}(BC^\ast) (1 - \mvh^2 - \sh)(\la - \uh^2)
  \right.
  \nonumber \\
  & & \left. \; \; \; \; \; \; \; \; \; \; \; \;
  + {\rm Re}(FG^\ast)
\left( (1 - \mvh^2 - \sh)(\la - \uh^2) + 4 \mlh^2 \la \right) \right]
        \Bigg.
        \nonumber \\
  & & \Bigg.
  - 2 \frac{\mlh^2}{\mvh^2} \la \left[
    {\rm Re}(FH^\ast)
    - {\rm Re}(GH^\ast) (1 - \mvh^2) \right]
 + |H|^2 \frac{\mlh^2}{\mvh^2} \sh \la
  \Bigg\} \; .
   \label{eq:ddwbvll}
\end{eqnarray}
Here the kinematic variables $(\sh,\uh)$ are defined as
\begin{eqnarray}
  \sh & = & \qh^2 = (\ph_+ + \ph_-)^2 \; , \\
  \uh & = & (\pbh - \ph_-)^2 - (\pbh - \ph_+)^2  \;
 \end{eqnarray}
which are bounded as
\begin{eqnarray}
  (2 \mlh)^2 \leq & \sh & \leq (1 - \hat{m}_{K,K^*})^2  \; ,
  \label{eq:sbound}\\
  -\uh(\sh) \leq & \uh & \leq \uh(\sh) \; ,
  \label{eq:ubound}
\end{eqnarray}
with $\ml=m_{\ell}/m_B$ and
\begin{eqnarray}
  \uh(\sh) & =& \sqrt{\la (1-4 \frac{\mlh^2}{\sh})}  \; , \\
 \la & =& 1+\hat{m}_{K,K*}^4+\sh^2-2 \sh-2 \hat{m}_{K,K*}^2(1+\sh) \;, \\
\D&=&\sqrt{1-\frac{4 \mlh^2}{s}}\;.
\end{eqnarray}
Note that the variable $\uh$ corresponds to $\theta$, the angle
between the momentum of the $B$-meson and the positively charged lepton
$\ell^+$  in the dilepton CMS frame, through the relation $\uh = -\uh(\sh) \cos \theta$
\cite{amm91}.

Integrating over $\uh$ in the kinematic region
given in Eq. (\ref{eq:ubound}) we get the formula of dilepton invariant mass
spectra (IMS)
\begin{eqnarray}
  \frac{\d \gl^K}{\d\sh} & = &
  \frac{G_F^2  \alpha^2  m_B^5}{2^{10} \pi^5}
      \left| V_{ts}^\ast  V_{tb} \right|^2  \uh(\sh)  D^{K}\\
D^{K}&=& (|\ap|^2 +|\cp|^2)
( \la- \frac{\uh(\sh)^2}{3} )+ |\s_1|^2 (\sh-4 \mlh^2) \nonumber \\
& & +  |\cp|^2 4 \ml^2 (2+2 \mph^2-\sh)
+ Re( \cp \ddp^{*}) 8 \ml^2 (1-\mph^2)
+|\ddp|^2 4 \ml^2 \sh \; ,\label{eq:dwbpll}\\
  \frac{\d \gl^{K^*}}{\d\sh} & = &
  \frac{G_F^2 \, \alpha^2 \, m_B^5}{2^{10} \pi^5}
      \left| V_{ts}^\ast  V_{tb} \right|^2 \, \uh(\sh) D^{K^*}\\
D^{K^*}  &= & \frac{|A|^2}{3} \sh \la (1+2 \frac{\mlh^2}{\sh})
+|E|^2 \sh \frac{\uh(\sh)^2}{3} + |\s_2|^2 (\sh-4 \mlh^2) \la
        \nonumber \\
  & & + \frac{1}{4 \mvh^2} \left[
|B|^2 (\la-\frac{\uh(\sh)^2}{3} + 8 \mvh^2 (\sh+ 2 \mlh^2) )
          + |F|^2 (\la -\frac{ \uh(\sh)^2}{3} + 8 \mvh^2 (\sh- 4 \mlh^2))
\right]       \nonumber \\
  & & +
   \frac{\la }{4 \mvh^2} \left[ |C|^2 (\la - \frac{\uh(\sh)^2}{3})
 + |G|^2 \left(\la -\frac{\uh(\sh)^2}{3}+4 \mlh^2(2+2 \mvh^2-\sh) \right)
\right]      \nonumber \\
  & & -  \frac{1}{2 \mvh^2}
\left[ {\rm Re}(BC^\ast) (\la -\frac{ \uh(\sh)^2}{3})(1 - \mvh^2 - \sh)
\nonumber  \right. \\
& & +    \left.   {\rm Re}(FG^\ast) ((\la -\frac{ \uh(\sh)^2}{3})(1 -
\mvh^2 - \sh) +  4 \mlh^2 \la) \right]
        \nonumber \\
  & & -  2 \frac{\mlh^2}{\mvh^2} \la  \left[ {\rm Re}(FH^\ast)-
 {\rm Re}(GH^\ast) (1-\mvh^2) \right] +\frac{\mlh^2}{\mvh^2} \sh \la |H|^2
 \; .
   \label{eq:dwbvll}
\end{eqnarray}

Both distributions agree with the ones obtained in \cite{abhh,gk},
if $C_{Q_1,2}$ are set to zero.

The differential FBA is defined as
$$ A_{FB} (s) = \frac{\displaystyle{- \int_0^{u(\sh)} dz \frac{d \Gamma}{ds du}+
\int_{-u(\sh)}^0du
\frac{d \Gamma}{ds du}}}{\displaystyle{\int_0^{u(\sh)} dz \frac{d \Gamma}{ds
du}+
\int_{-u(\sh)}^0du\frac{d \Gamma}{ds du}}}~.$$
For $B\to K\ell^+\ell^-$ decays it reads as follows
\begin{eqnarray}
  \frac{\d \a_{\rm FB}^{K}}{\d \sh} D^K& =& - 2 \mlh \uh(\sh){\rm Re}(\s_1 \ap^*)
\label{eqn:bkas}
\end{eqnarray}
For $B\to K^*\ell^+\ell^-$ decays it reads as follows
\begin{eqnarray}
  \frac{\d \a_{\rm FB}^{K^*}}{\d \sh} D^{K^*}& =& \uh(\sh)
      \Bigg \{ \sh \left [ {\rm Re}(B E^\ast) + {\rm Re}(AF^\ast)\right ]
\Bigg. \; \nonumber \\
&&\Bigg. +\frac{\mlh}{\mvh} \left [{\rm Re}(\s_2
B^*) (1-\sh-\mvh^2) -{\rm Re}(\s_2 C^*)\la \right ]
\Bigg \}
\label{eqn:bksas}
\end{eqnarray}

We can read from (\ref{eqn:bkas}), the FBA  of the process $B\rightarrow K
\ell^+ \ell^-$ does not vanish when the contributions of NHB are taken into
account. With it, our analysis below also show that the contributions of NHBs can even be accessible
in B factories.

The lepton polarization can be defined as follows
\ba
\frac{d \Gamma ( \vec{n} )}{d s} &=&
\frac{1}{2} ( \frac{d \Gamma}{d s} )_0
\Big[ 1 + ( P_L\, \vec{e}_L + P_N\, \vec{e}_N + P_T\, \vec{e}_T ) \cdot
\vec{n} \Big]~,
\ea
where the subscript "0" corresponds to the unpolarized width, and
$P_L$, $P_T$, and $P_N$, correspond to the longitudinal, transverse and
normal components of the polarization vector, respectively.

For the process $B\rightarrow K \ell^- \ell^+$, the $P_L^K$, $P_T^K$, and
$P_N^K$, are derived respectively as
\ba
P_L^K D^K&=&\frac{4}{3} \D \Bigg \{ \la {\rm Re}(\ap \cp^{\ast})-3 \mlh
(1-\mph^2) {\rm Re}(\cp^{\ast} \s_1) -3 \mlh \sh {\rm Re}(\rp^*
\s_1) \Bigg \}, \label{eqn:bklp} \\
P_N^K D^K&=&\frac{\pi \sqrt{\sh} \uh(\sh)}{2} \Bigg \{-{\rm Im}(\ap \s_{1}^*)+ 2 \mlh {\rm Im}(\cp \rp^*)\Bigg \},\\
P_T^K D^K&=&\frac{-\pi \sqrt{\la}}{\sqrt{\sh}} \Bigg \{\mlh \left [(1-\mph^2) {\rm
Re}(\ap \cp^*)\Bigg. \right. \nonumber \\
&&\Bigg.\left. +\sh {\rm Re}(\ap \rp^{\ast})\right ] + \frac{(\sh-4\mlh^2)}{2}{\rm
Re}(\cp \s_{1}^{\ast}) \Bigg \}.      \label{eqn:bktp}\ea

$D^K$ is defined in Eq. (\ref{eq:dwbpll}). For the process $B\rightarrow K^* \ell^- \ell^+$, the $P_L^{K^*}$,
$P_T^{K^*}$, and $P_N^{K^*}$, are derived respectively as
\ba
P_L^{K^*} D^{K^*}&=&\D \Bigg \{\frac{2 \sh \la}{3} {\rm Re}(A E^*) +
\frac{(\la+12 \mvh^2)}{3 \mvh^2}{\rm Re}(B F^*) \Bigg. \nonumber\\
&&\Bigg.-\frac{\la (1-\mvh^2-\sh)}{3
\mvh^2}{\rm Re}(B G^*+C F^*) +\frac{\la^2}{3 \mvh^2} {\rm Re}
(C G^*)\Bigg. \nonumber \\
&&\Bigg. +\frac{2 \mlh \la}{\mvh} \left[{\rm Re}(F \s_2^*) - \sh {\rm Re}(H
\s_2^*)-(1-\mvh^2){\rm Re}(G \s_2^*)\right ] \Bigg \},   \label{eqn:bkslp} \\
P_N^{K^*} D^{K^*}&=&\frac{- \pi \sqrt{\sh} \uh(\sh)}{4 \mph}
\Bigg \{\frac{\mlh}{\mvh}\left [ {\rm Im}(F G^*)(1+3 \mvh^2-s)\right.\Bigg.\nonumber \\
&&\left.\Bigg.+{\rm Im}(F H^*)(1-\mvh^2-s)-{\rm Im}(G H^*)\la\right ]\Bigg.\nonumber \\
&&\Bigg.+2 \mvh \mlh
\left [{\rm Im} (B E^*)+{\rm Im}(A F^*)\right ] \Bigg.\nonumber \\
&&\Bigg. -(1-\mvh^2-\sh) {\rm Im} (B \s_2^*)+\la {\rm Im}(C
\s_2^*)\Bigg \}, \\
P_T^{K^*} D^{K^*}&=&\frac{\pi \sqrt{\la} \mlh}{4 \sqrt{\sh}}\Bigg \{ 4 \sh {\rm Re}(A B^*) \Bigg.\nonumber \\
&&\Bigg.+\frac{(1-\mvh^2 -\sh)}{\mvh^2}\left[-{\rm Re}(B F^*)+(1-\mvh^2){\rm Re}(B G^*)+\sh{\rm
Re}(B H^*)\right ]\Bigg.
\nonumber \\
&&\Bigg.+\frac{\la}{\mvh^2} \left [{\rm Re} (C F^*) - (1-\mvh^2){\rm Re}(C G^*)-\sh
{\rm Re}(C H^*) \right ] \Bigg.\nonumber \\
&&\Bigg.+\frac{(\sh-4 \mlh^2)}{\mvh \mlh} \left[(1-\mvh^2-\sh){\rm Re} (F \s_2^*)
-\la {\rm Re}(G \s_2^*)\right]\Bigg \}.
\label{eqn:bkstp}
\ea
$D^{K^*}$ is defined by Eq.(\ref{eq:dwbvll}).
\section{numerical analysis}
Parameters used in our analysis are list in Table \ref{tab:para}. Considering that the branching ratios of
$\bkll$ and $\bksll$ are not very sensitive to the mass of $m_b$, we neglect the difference between the pole mass
and running mass of b quark.
\begin{table}
        \begin{center}
        \begin{tabular}{|l|l|}
        \hline
        $m_b$                   & $4.8$ GeV \\
        $m_c$                   & $1.4$ GeV \\
        $m_s$                   & $0.2$ GeV   \\
        $m_{\mu}$                & $0.11$ GeV \\
        $m_{\tau}$               & $1.78$ GeV \\
        $M_B$                   & $5.28$ GeV \\
        $M_K$                   & $0.49$ GeV \\
        $M_{K^*}$                   & $0.89$ GeV \\
        $M_{J/\psi}(M_{\psi'})$      & $3.10(3.69)$   GeV\\
        $\Gamma_B$                   & $4.22\times10^{-13}$ GeV \\
        $\Gamma_{J/\psi}(\Gamma_{\psi'})$      & $8.70(27.70)\times 10^{-5}$   GeV\\
        $\Gamma(J/\psi\rightarrow\ell^+\ell^-)$      & $5.26\times 10^{-6}$   GeV\\
        $\Gamma(\psi'\rightarrow\ell^+\ell^-)$      & $2.14\times 10^{-6}$   GeV\\
        $G_F$                         & $1.17\times 10^{-5}$ GeV$^{-2}$\\
        $\alpha^{-1}$     & 129           \\
        $|V^\ast_{ts} V_{tb}|$ & 0.0385 \\
        \hline
        \end{tabular}
        \end{center}
\caption{\it Values of the input parameters used in our numerical analysis.}
\label{tab:para}
\end{table}

The Wilson coefficients in the SM used in the numerical analysis is given in the Table \ref{tab:wcsm}.
$C_7^{eff}$ is defined as
\ba
C_7^{eff} &=& C_7 -C_5/3 -C_6\;.
\ea
\begin{table}
        \begin{center}
        \begin{tabular}{|c|c|c|c|c|c|c|c|c|c|}
        \hline
        \multicolumn{1}{|c|}{ $C_1$}       &
        \multicolumn{1}{|c|}{ $C_2$}       &
        \multicolumn{1}{|c|}{ $C_3$}       &
        \multicolumn{1}{|c|}{ $C_4$}       &
        \multicolumn{1}{|c|}{ $C_5$}       &
        \multicolumn{1}{|c|}{ $C_6$}       &
        \multicolumn{1}{|c|}{ $C_7^{\rm eff}$}       &
        \multicolumn{1}{|c|}{ $C_9$}       &
        \multicolumn{1}{|c|}{$C_{10}$} &
        \multicolumn{1}{|c|}{ $C$ }     \\
        \hline
        $-0.248$ & $+1.107$ & $+0.011$ & $-0.026$ & $+0.007$ & $-0.031$ &
   $-0.313$ &   $+4.344$ &    $-4.669$    & $+0.362$     \\
        \hline
        \end{tabular}
        \end{center}
\caption{ \it Wilson coefficients of the SM used in the numerical
          analysis.}
\label{tab:wcsm}
\end{table}

$C_{Q_{1,2}}$ come from  exchanging  NHBs and are proportional to tan$^3\beta$ in some regions of the
parameter space in SUSY models.
According to the analysis in \cite{hly,hllyz}, the necessary conditions for the large contributions of NHBs include: ({\it i})
the ratio of vacuum expectation value, tan$\beta$, should be large, ({\it ii}) the mass values of the lighter
chargino and the lighter stop should not be too large (say less than 120 GeV), ({\it iii})mass splitting of charginos
and stops should be large, which also indicate large mixing between stop sector and chargino sector.
As the conditions are satisfied, the process $B\rightarrow X_s \gamma$ will impose a constraint on $C_7$.
It is well known that this process puts a very stringent constraint on the possible new physics and that
SUSY can contribute destructively when the signature of the Higgs mass term $\mu$ is minus.
There exist considerable regions of SUSY parameter space in which NHBs can largely contribute
to the process $b\rightarrow s \ell^+ \ell^-$ while the constraint of $b\rightarrow s \gamma$ is respected
(i.e., the signature of the Wilson coefficient $C_7$ is changed from positive to negative).
When the masses of SUSY particle are relatively heavy (say, 450 Gev),
there are still significant regions in the parameter space of SUSY models in which NHBs could contribute largely.
However, at these cases $C_7$ does not change its sign, because contributions of charged Higgs and
charginos cancel with each other. We will see it is hopeful to distinguish these two kinds of regions of
SUSY parameter space through observing $B\rightarrow K^{(*)}\ell^+\ell^-$.

As pointed out in \cite{dhh,hly}, the contribution of NHBs is proportional to the lepton mass, therefore for $\ell=e$,
contributions of NHBs can be safely neglected. While for cases $\ell=\mu$ and $\ell=\tau$, the contributions of NHBs
can be considerably large. To investigate the effects of NHBs in SUSY models, we take typical
values of $C_{7,9,10}$ and  $C_{Q_{1,2}}$ as given in Table \ref{tab:susy}. The SUSY model without
considering the effects of NHBs (SUSY I in Table 4) is given as a reference frame so that could
the effects of NHBs be shown in high relief.
\begin{table}
        \begin{center}
        \begin{tabular}{|c|c|c|c|c|c|}
        \hline
        \multicolumn{1}{|c|}{ SUSY models} &
        \multicolumn{1}{c|}{ $R_7$}       &
        \multicolumn{1}{c|}{ $R_9$}       &
        \multicolumn{1}{c|}{ $R_{10}$}       &
        \multicolumn{1}{c|}{ $C_{Q_1}$}       &
        \multicolumn{1}{c|}{ $C_{Q_2}$}      \\
        \hline
        SUSY I  & $-1.2$ & $1.1$ & $0.8$ & $0.0$ & $0.0$ \\
        \hline
        SUSY II  & $-1.2$ & $1.1$ & $0.8$ & $6.5(16.5)$ & $-6.5(-16.5)$ \\
        \hline
        SUSY III  & $1.2$ & $1.1$ & $0.8$ & $1.2(4.5)$ & $-1.2(-4.5)$ \\
        \hline
        \end{tabular}
        \end{center}
\caption{ \it Wilson coefficients of the SUSY used in our numerical
        analysis. $R_i$ means $C_i$/$C_i^{SM}$. SUSY I corresponds to the regions where SUSY can destructively
        contribute and can change the sign of $C_7$, but the contributions of NHBs are neglected.
        SUSY II corresponds to the regions where tan$\beta$ is large and the masses of superpartners
        are relatively small. SUSY III corresponds to the regions where tan$\beta$ is large but the masses of
        superpartners are relatively large. In the last two cases the effects of NHBs are taken into account.
        The contributions of NHBs are settled to be different for both the case $\ell=\mu$ and $\ell=\tau$, since $C_{Q_{1,2}}$
        are proportional to the mass of lepton. The values in bracket are for the case $\ell=\tau$.} \footnote{The reference \cite{ln} points out that in 2HDM, the missed box diagram can contribute considerably. But in SUSY, the dominant contribution is from the SUSY self-energy diagram which is proportional to tan$^3\beta$, therefore the missed box diagram do not modify the result of SUSY.}
\label{tab:susy}
\end{table}
\begin{figure}
\begin{minipage}[t]{8.2cm}
     \epsfig{file=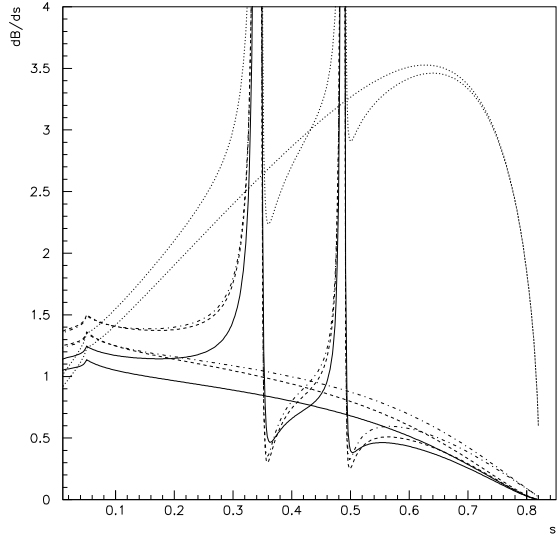,width=8.2cm}
     \mbox{ }\hfill\hspace{1cm}(a)\hfill\mbox{ }
     \end{minipage}
     \hspace{-0.4cm}
     \begin{minipage}[t]{8.2cm}
     \epsfig{file=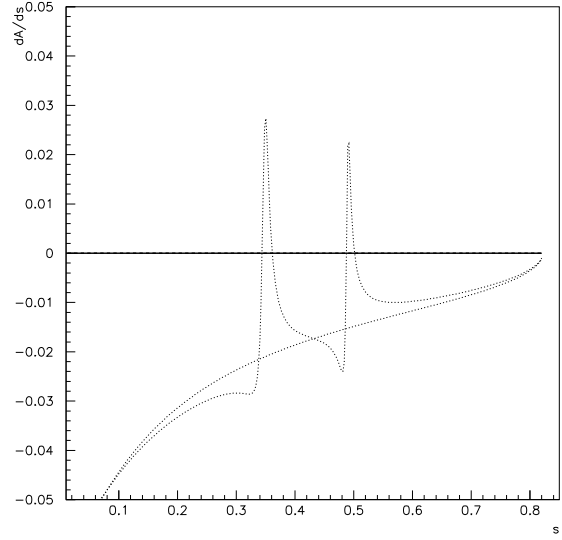,width=8.2cm}
     \mbox{ }\hfill\hspace{1cm}(b)\hfill\mbox{ }
     \end{minipage}
\vskip 0.05truein
     \begin{minipage}[t]{8.2cm}
     \epsfig{file=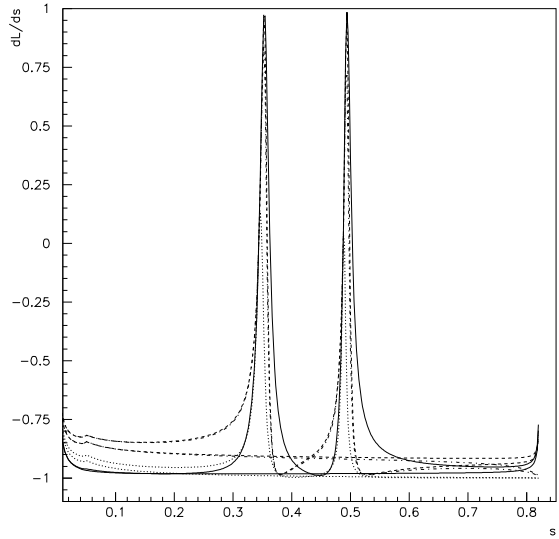,width=8.2cm}
     \mbox{ }\hfill\hspace{1cm}(c)\hfill\mbox{ }
     \end{minipage}
     \hspace{-0.4cm}
     \begin{minipage}[t]{8.2cm}
     \epsfig{file=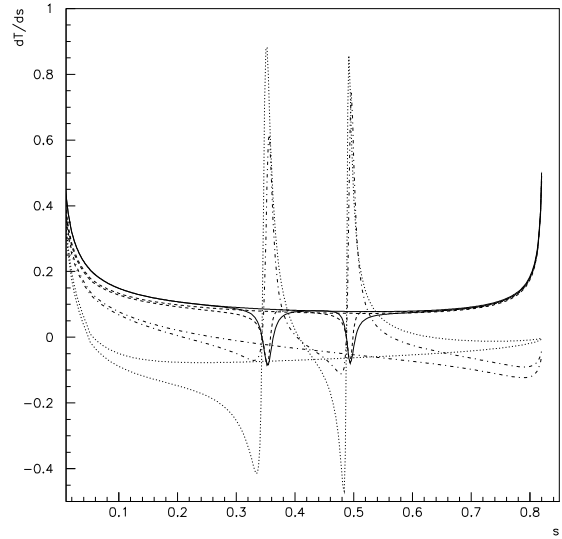,width=8.2cm}
     \mbox{ }\hfill\hspace{1cm}(d)\hfill\mbox{ }
     \end{minipage}
     \caption{\it The IMS(a), FBA(b), $P_L$(c), and $P_T$(d) of the process $\bkm$. The solid line, dashed line, dot line and dashed-dot line represent
   the SM, SUSY I, SUSY II, SUSY III respectively. Both the total (SD+LD) and the pure SD contributions are shown in order to compare.}
\label{fig:kman}
\end{figure}

Numerical results are shown in Figs. 1-4.
In Fig. \ref{fig:kman}(a), the IMS of $B\rightarrow K\mu^+\mu^-$ is depicted. We see that
at the high $\sh$ regions, NHBs greatly modify the spectrum. While at the low $\sh$ region, the effects of NHBs
become weak. In Fig. \ref{fig:kman}(b), the FBA of the $B\rightarrow K \mu^+ \mu^-$ is presented.
Fig. \ref{fig:kman}(b) shows that the average FBA in $B\rightarrow K \mu^+ \mu^-$ is 0.02. To measure an asymmetry $A$ of a decay with the
branching ratio $Br$ at the $n\sigma$ level, the required number of events is $N=n^2/(Br A^2)$. For $B \rightarrow K\mu^+ \mu^-$,
the average FBA is 0.02 or so, the required number of events is $10^{12}$ or so. Therefore it is hard to observe
the derivation of FBA from the SM.
In Fig. \ref{fig:kman}(c) and Fig. \ref{fig:kman}(d), the longitudinal and transverse polarizations are given.
The effect of NHBs to the longitudinal polarization is weak but the effect to the transverse is remarkable.
\begin{figure}
\begin{minipage}[t]{8.2cm}
     \epsfig{file=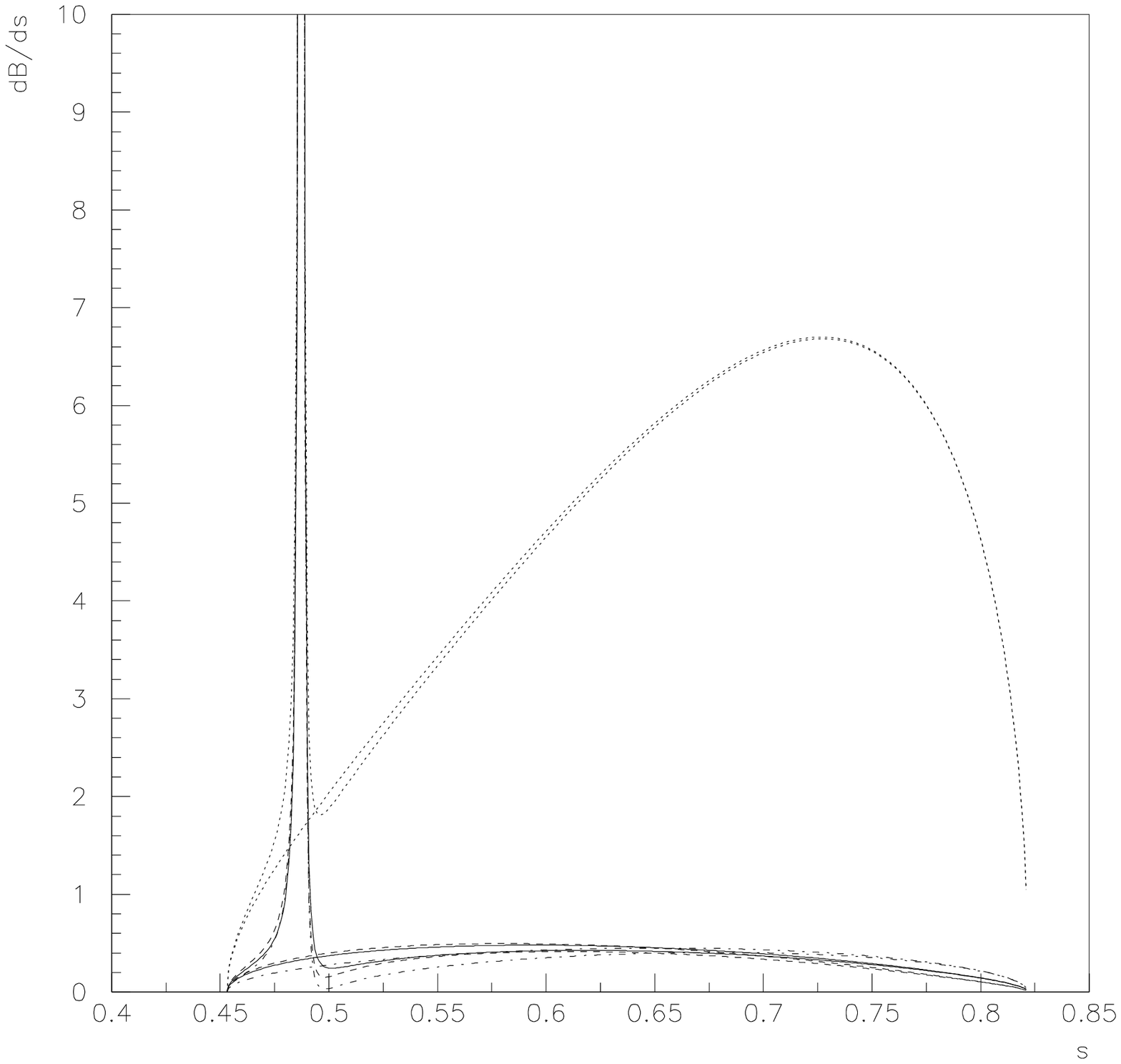,width=8.2cm}
     \mbox{ }\hfill\hspace{1cm}(a)\hfill\mbox{ }
     \end{minipage}
     \hspace{-0.4cm}
     \begin{minipage}[t]{8.2cm}
     \epsfig{file=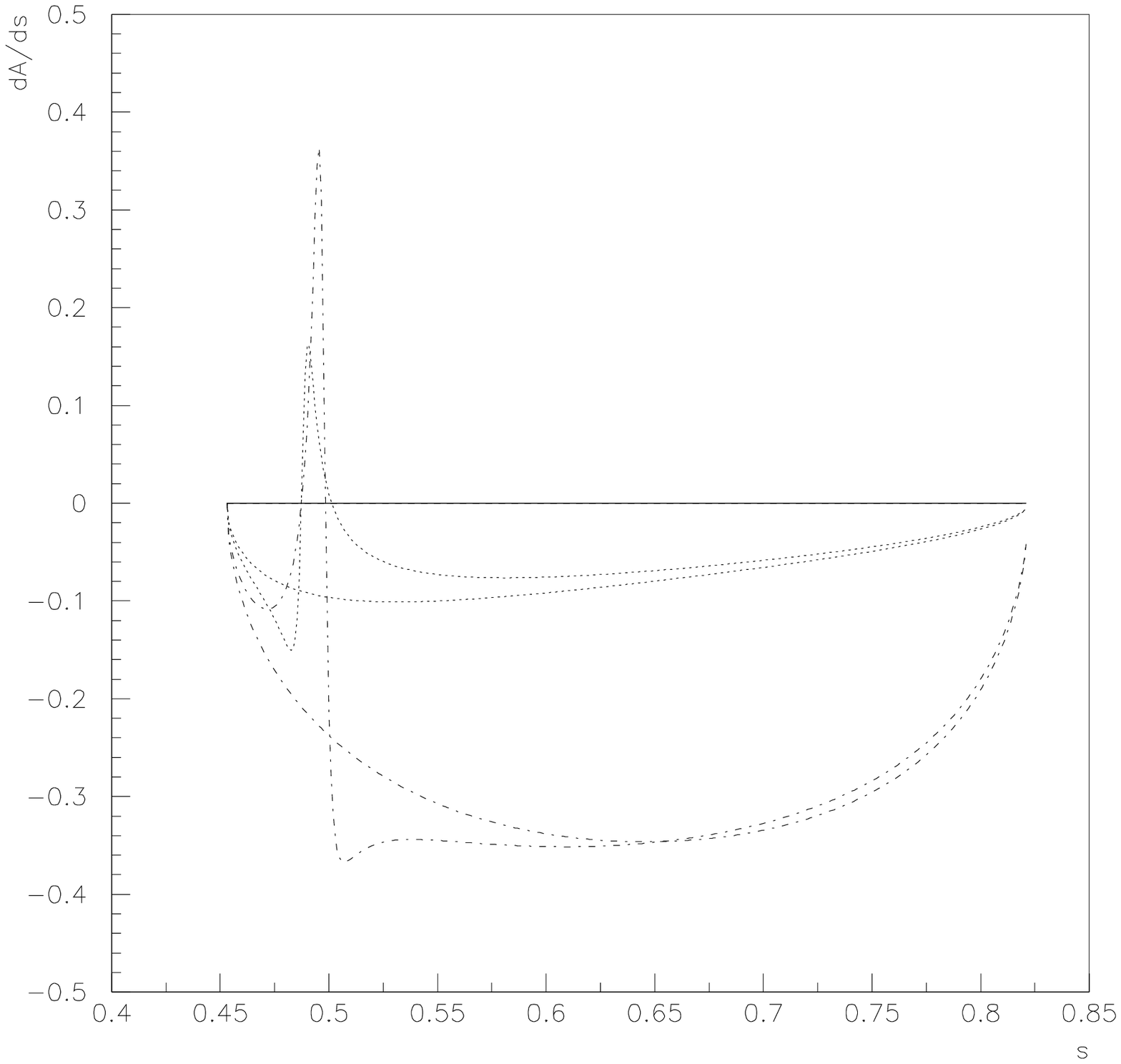,width=8.2cm}
     \mbox{ }\hfill\hspace{1cm}(b)\hfill\mbox{ }
     \end{minipage}
\vskip 0.05truein
     \begin{minipage}[t]{8.2cm}
     \epsfig{file=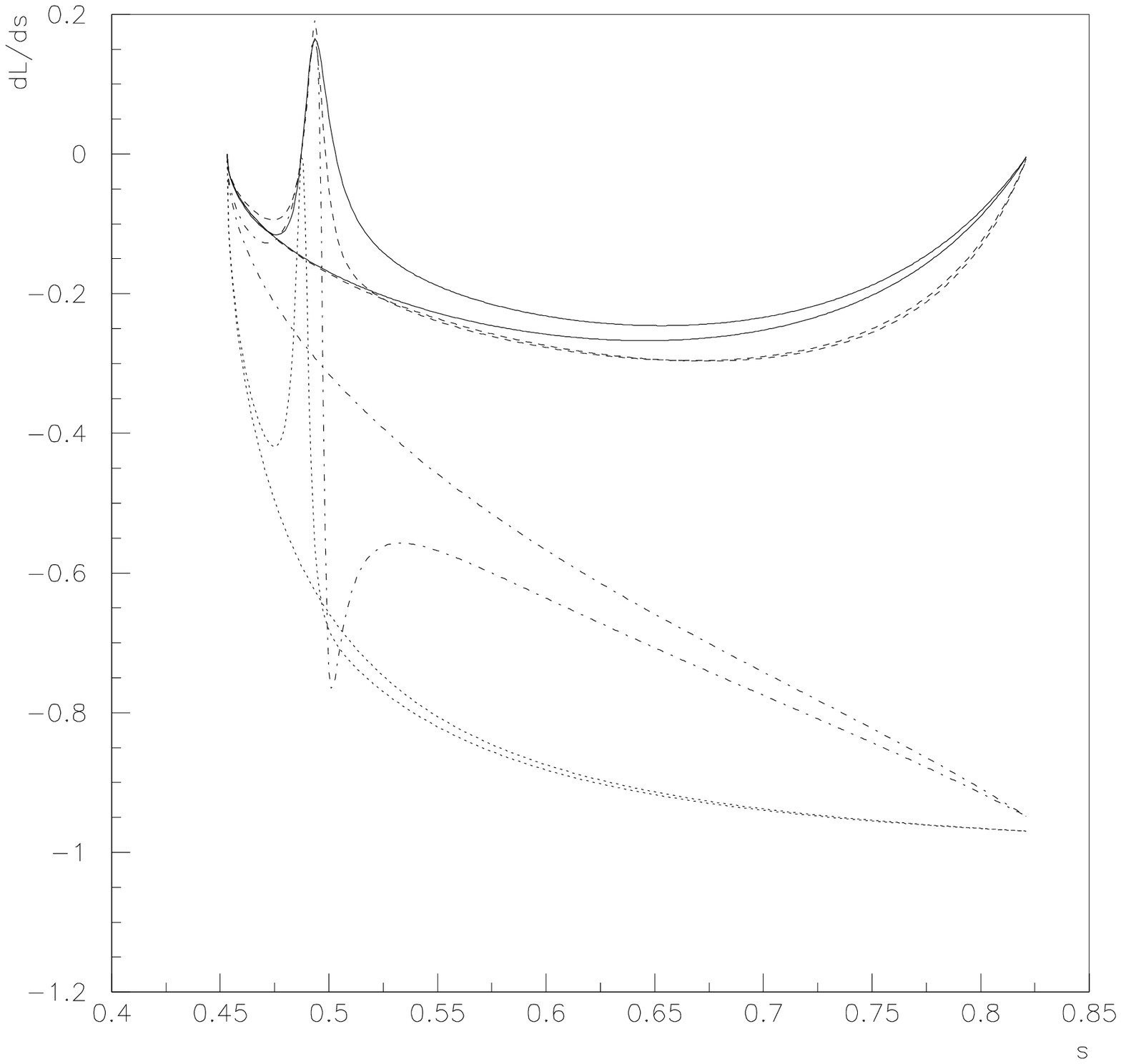,width=8.2cm}
     \mbox{ }\hfill\hspace{1cm}(c)\hfill\mbox{ }
     \end{minipage}
     \hspace{-0.4cm}
     \begin{minipage}[t]{8.2cm}
     \epsfig{file=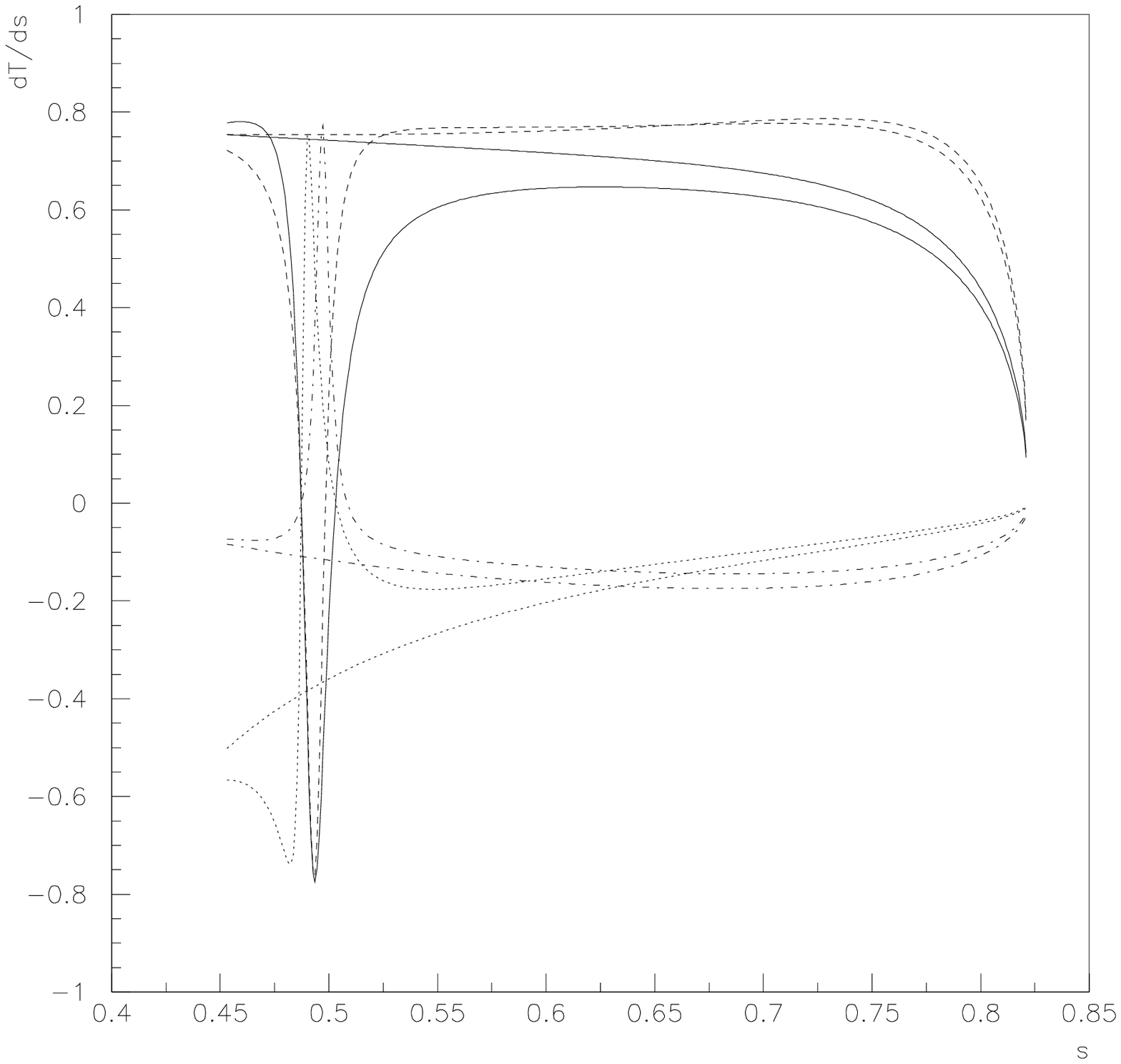,width=8.2cm}
     \mbox{ }\hfill\hspace{1cm}(d)\hfill\mbox{ }
     \end{minipage}
     \caption{\it The IMS(a), FBA(b), $P_L$(c), and $P_T$(d) of the process $\bkt$. The line conventions are the
same as given in the legend of Fig 1.}
\label{fig:ktan}
\end{figure}

In Fig. \ref{fig:ktan}(a) and Fig. \ref{fig:ktan}(b) the IMS and FBA  of $B\rightarrow K\tau^+\tau^-$ are presented respectively.
For SUSY II, the effects of NHBs to IMS is quite manifest, and the average FBA can reach 0.1.
For SUSY III, the average FBA can reach $0.3$. Therefore, in order to observe FBA, the required number
of events should be $10^{9}$ or so and  $10^{8}$, respectively, so that in B factories, say LHCB, these two
cases are accessible.
In Fig. \ref{fig:ktan}(c) and Fig. \ref{fig:ktan}(d), the longitudinal and transverse polarizations are drawn respectively. The effects of
NHBs are also very obvious.

\begin{figure}
\begin{minipage}[t]{8.2cm}
     \epsfig{file=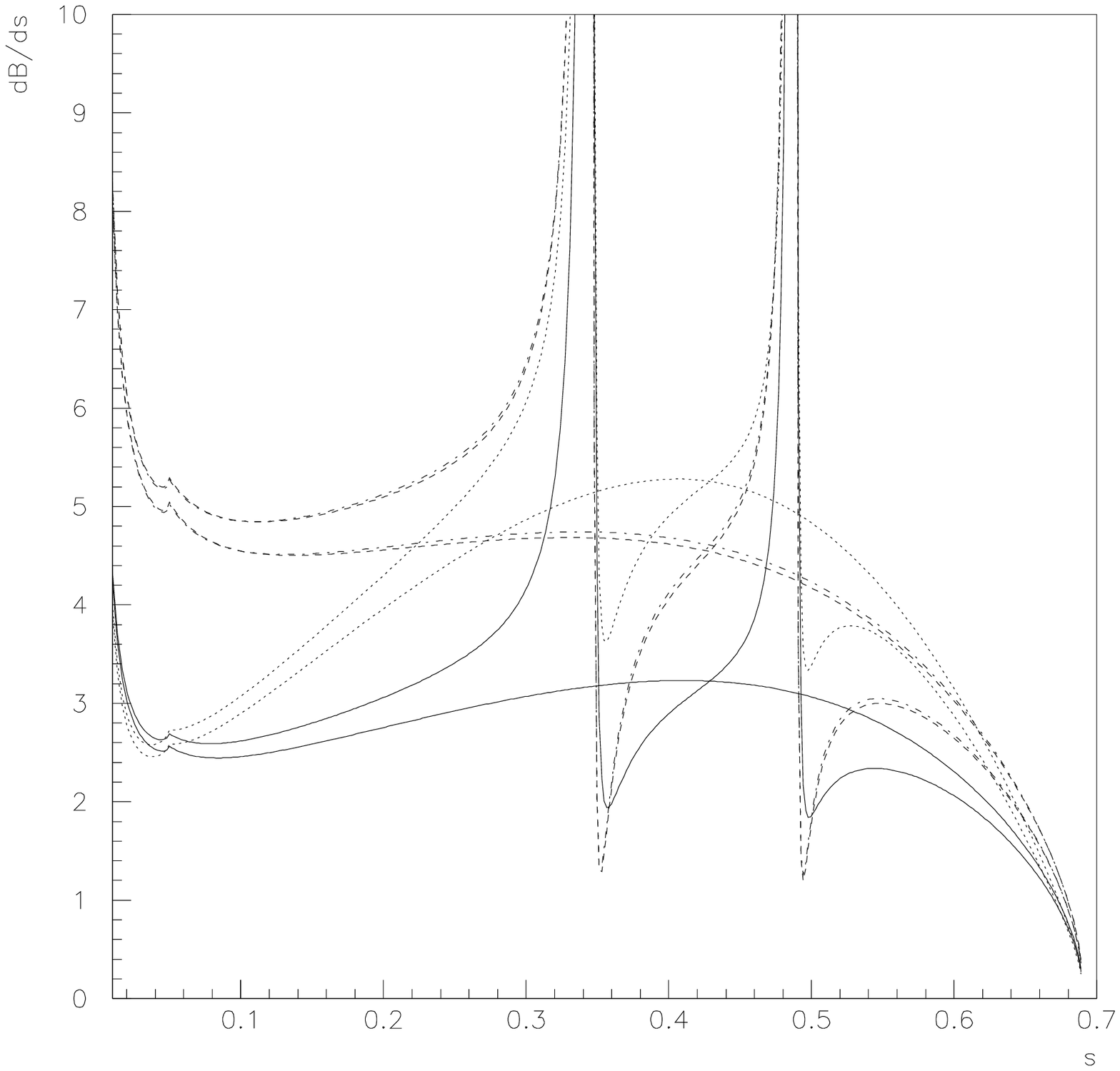,width=8.2cm}
     \mbox{ }\hfill\hspace{1cm}(a)\hfill\mbox{ }
     \end{minipage}
     \hspace{-0.4cm}
     \begin{minipage}[t]{8.2cm}
     \epsfig{file=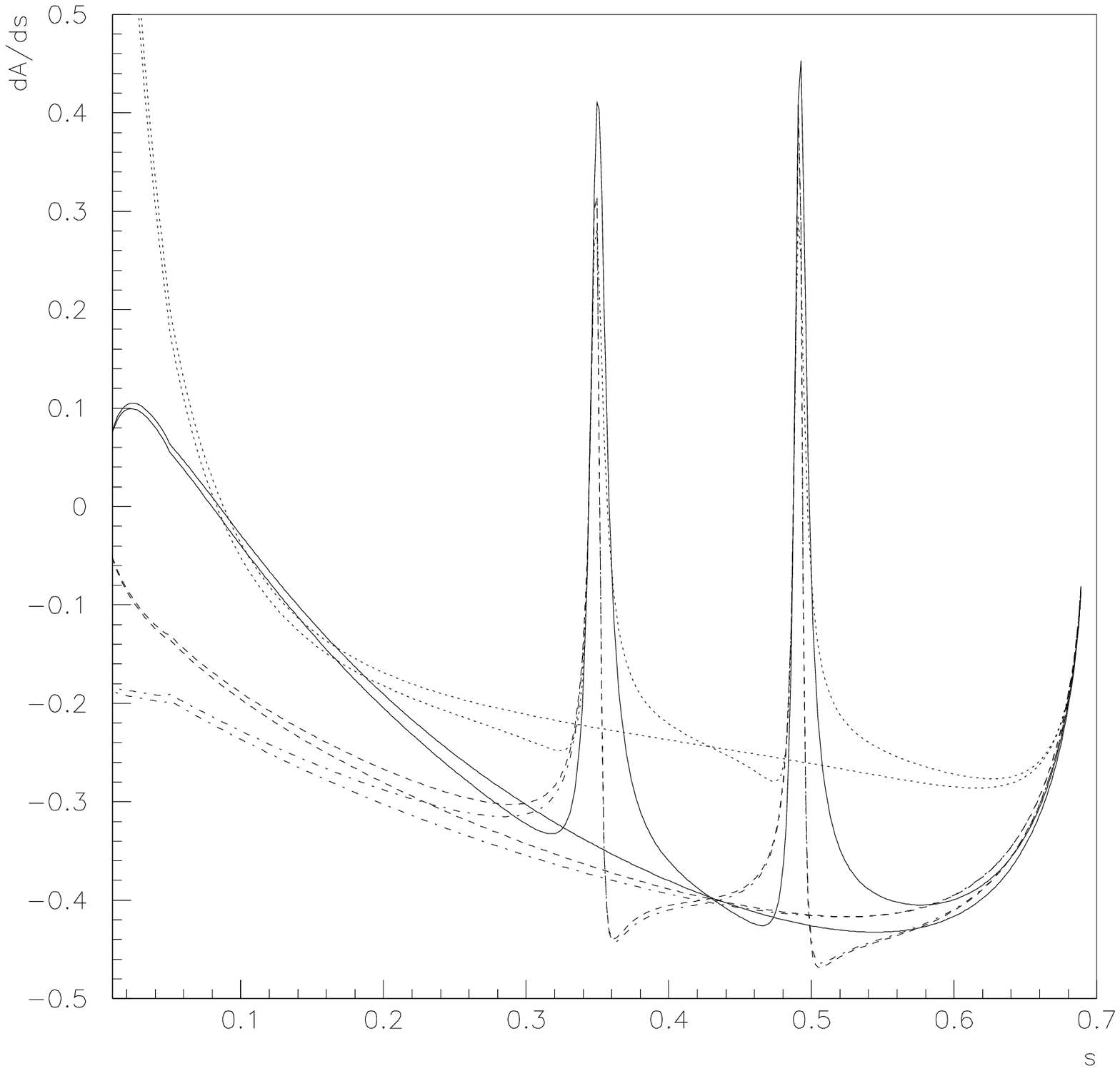,width=8.2cm}
     \mbox{ }\hfill\hspace{1cm}(b)\hfill\mbox{ }
     \end{minipage}
\vskip 0.05truein
     \begin{minipage}[t]{8.2cm}
     \epsfig{file=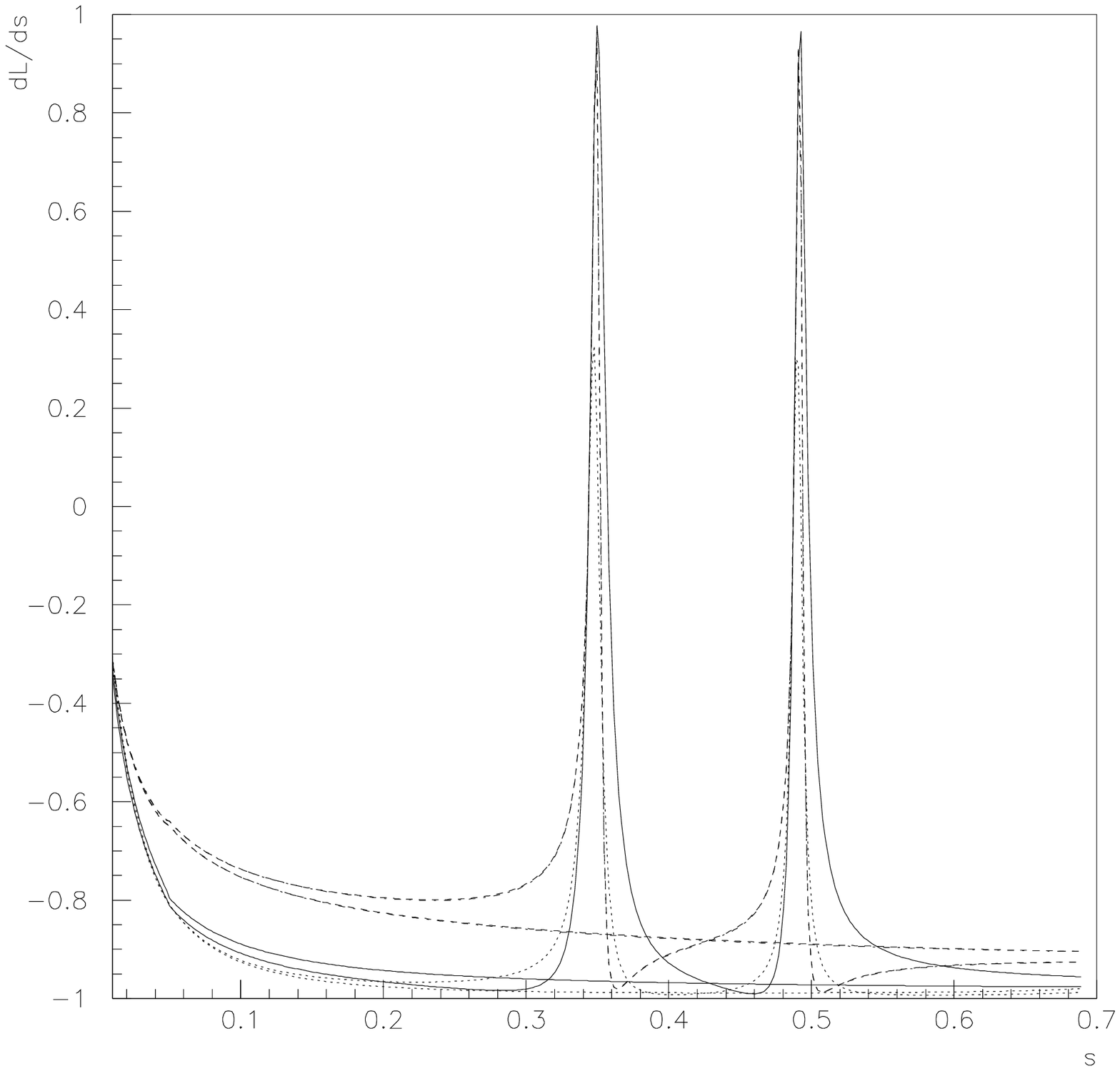,width=8.2cm}
     \mbox{ }\hfill\hspace{1cm}(c)\hfill\mbox{ }
     \end{minipage}
     \hspace{-0.4cm}
     \begin{minipage}[t]{8.2cm}
     \epsfig{file=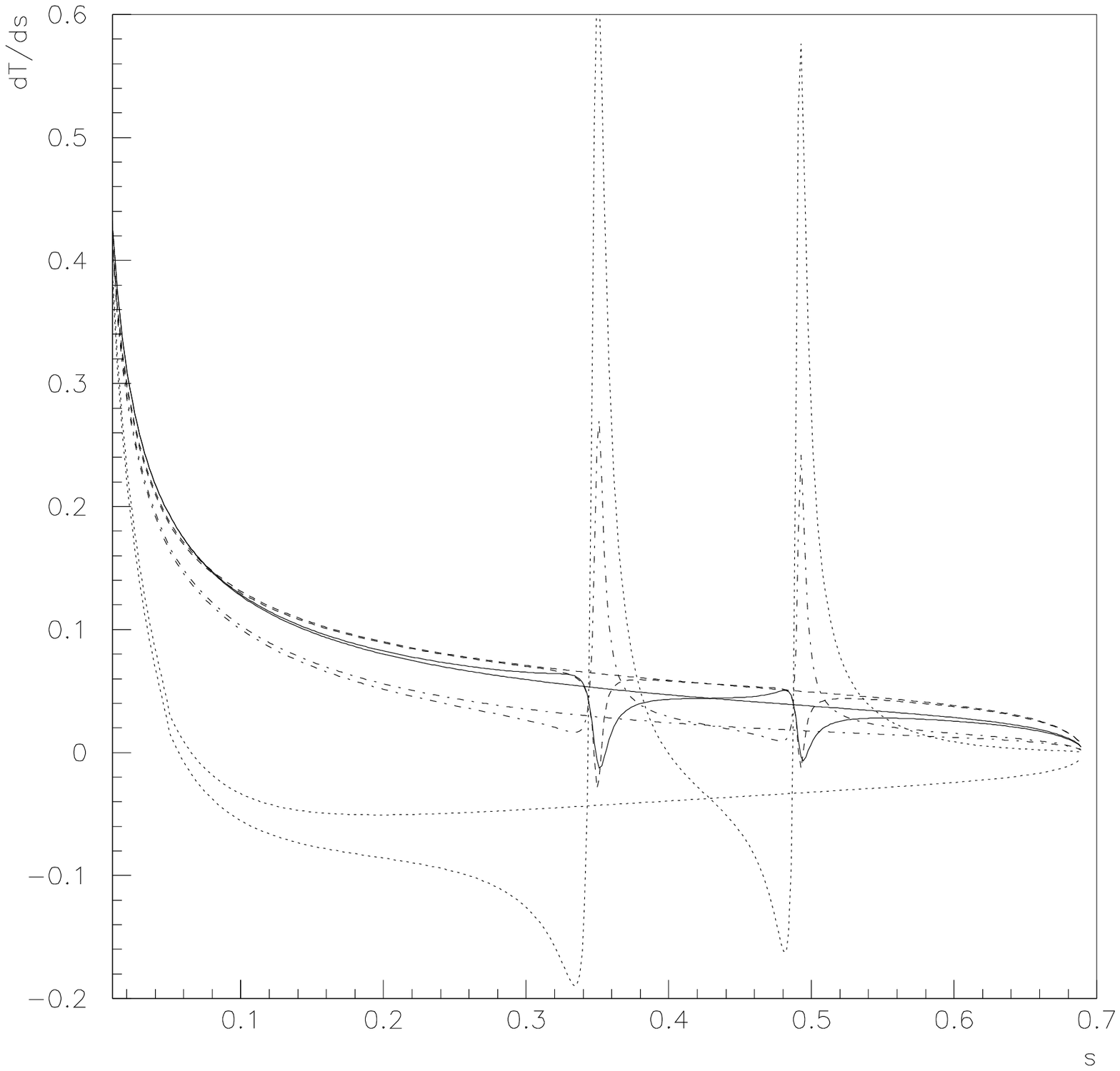,width=8.2cm}
     \mbox{ }\hfill\hspace{1cm}(d)\hfill\mbox{ }
     \end{minipage}
     \caption{\it The IMS(a), FBA(b), $P_L$(c), and $P_T$(d) of the process $\bksm$. The line conventions are the
same as given in the legend of Fig 1.}
\label{fig:ksman}
\end{figure}

Figs. 3 and 4 are devoted to the decay $B\rightarrow K^{*} l^+l^-$.
In Fig. \ref{fig:ksman}, the IMS, FBA, and polarizations of $B\rightarrow K^* \mu^+ \mu^-$ are given. We see that
this process is not as much as sensitive to the effect of NHB as $\bkm$. However, the contribution of NHBs
will increase the part with positive FBA and will be helpful to determine the zero
point of FBA.  Fig. \ref{fig:ksman}(d) depicts
the transverse polarization of the $\bksm$, and the effect of NHBs is quite obvious.
The zero point of the FBA can be
slightly modified as shown in Figure \ref{fig:ksman}(b) due to the contributions of NHBs.

\begin{figure}
\begin{minipage}[t]{8.2cm}
     \epsfig{file=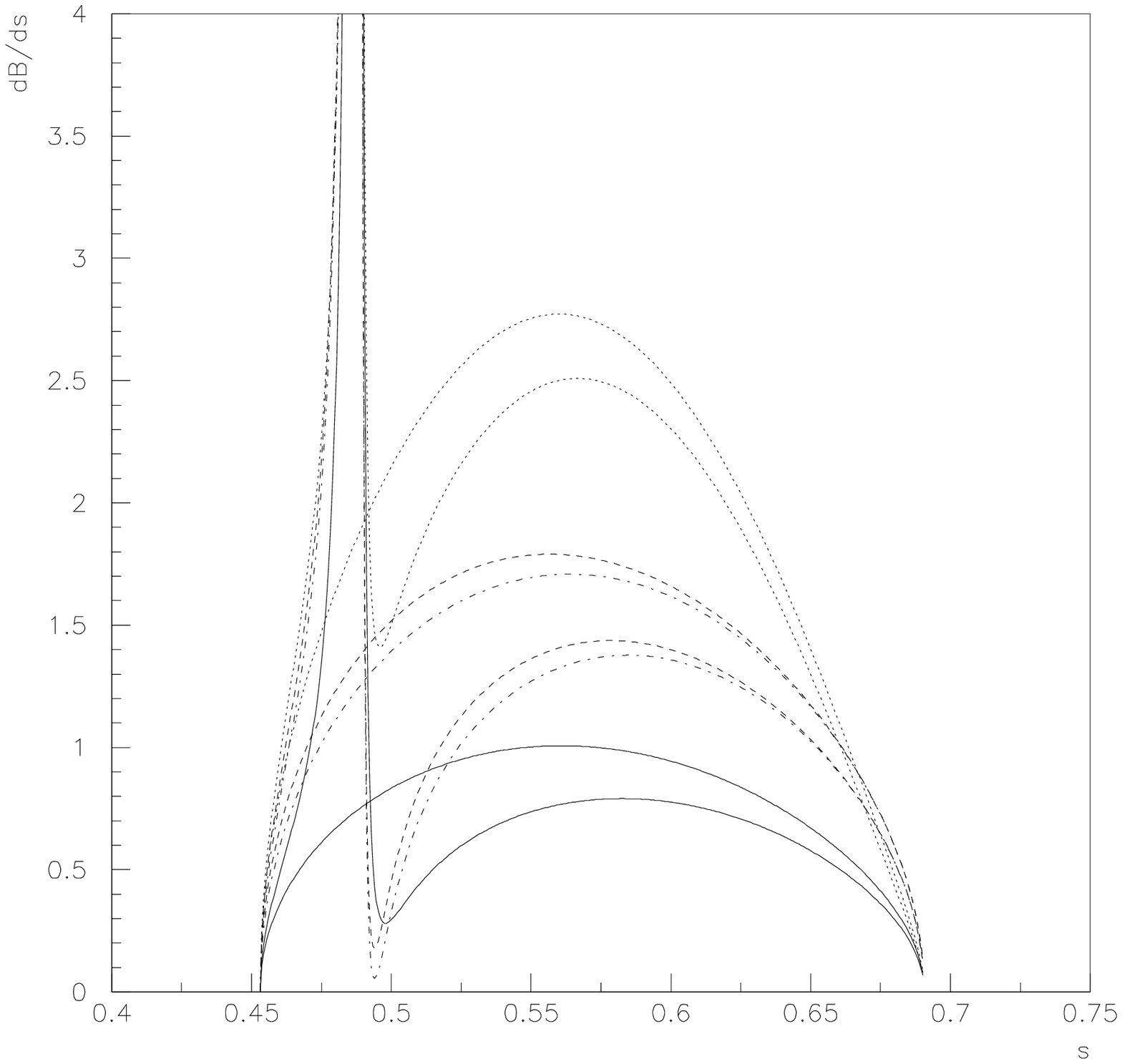,width=8.2cm}
     \mbox{ }\hfill\hspace{1cm}(a)\hfill\mbox{ }
     \end{minipage}
     \hspace{-0.4cm}
     \begin{minipage}[t]{8.2cm}
     \epsfig{file=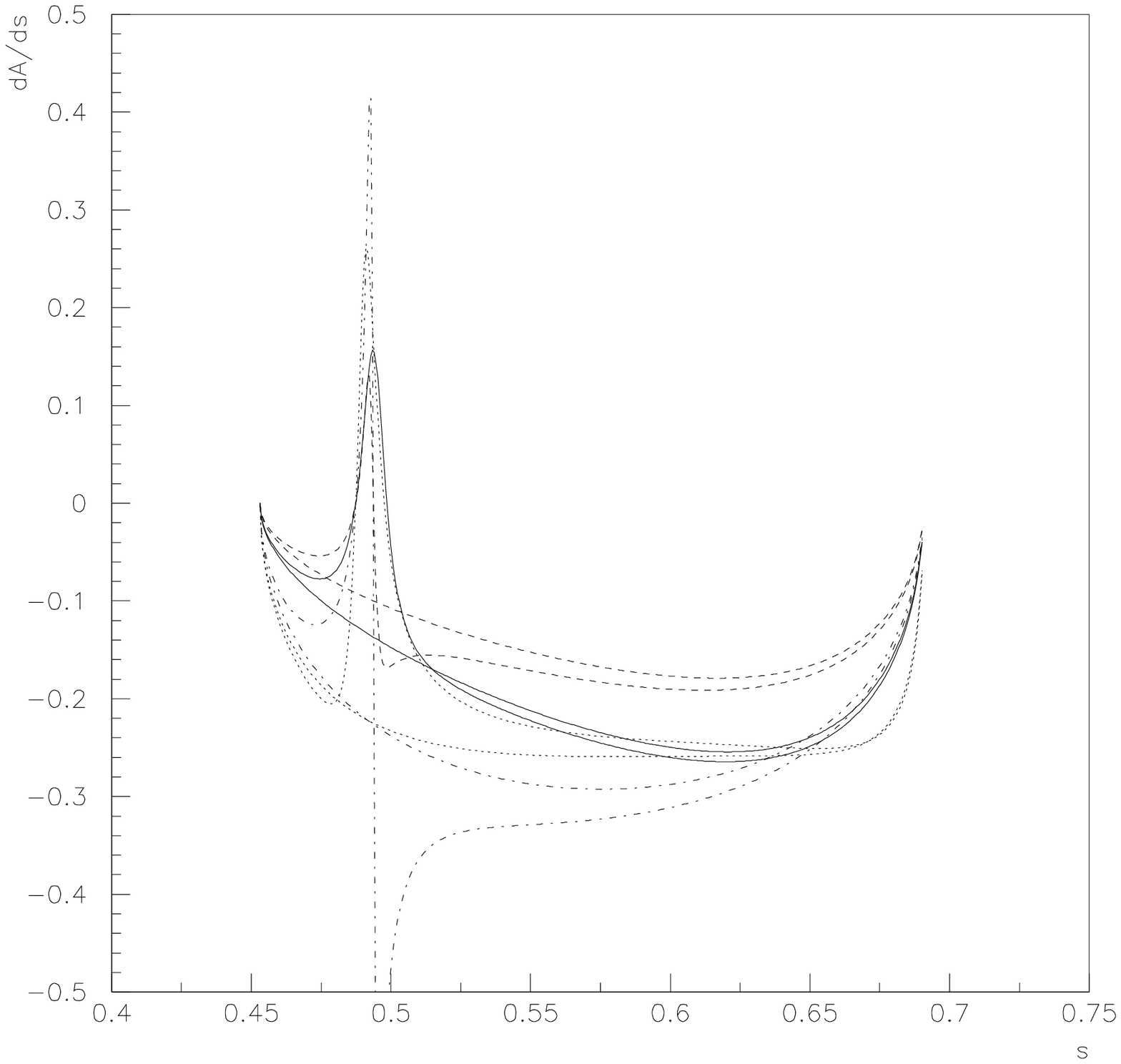,width=8.2cm}
     \mbox{ }\hfill\hspace{1cm}(b)\hfill\mbox{ }
     \end{minipage}
\vskip 0.05truein
     \begin{minipage}[t]{8.2cm}
     \epsfig{file=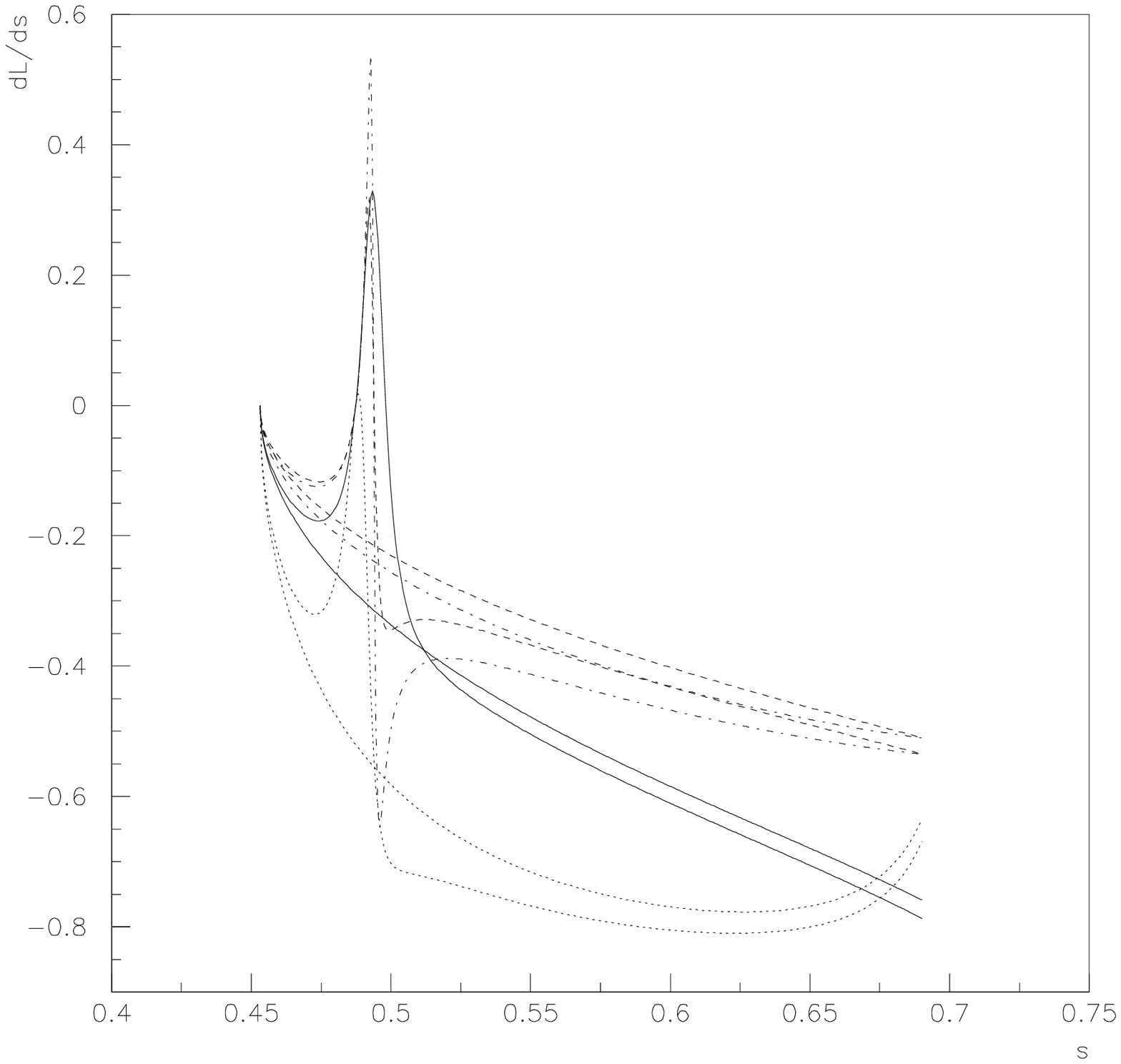,width=8.2cm}
     \mbox{ }\hfill\hspace{1cm}(c)\hfill\mbox{ }
     \end{minipage}
     \hspace{-0.4cm}
     \begin{minipage}[t]{8.2cm}
     \epsfig{file=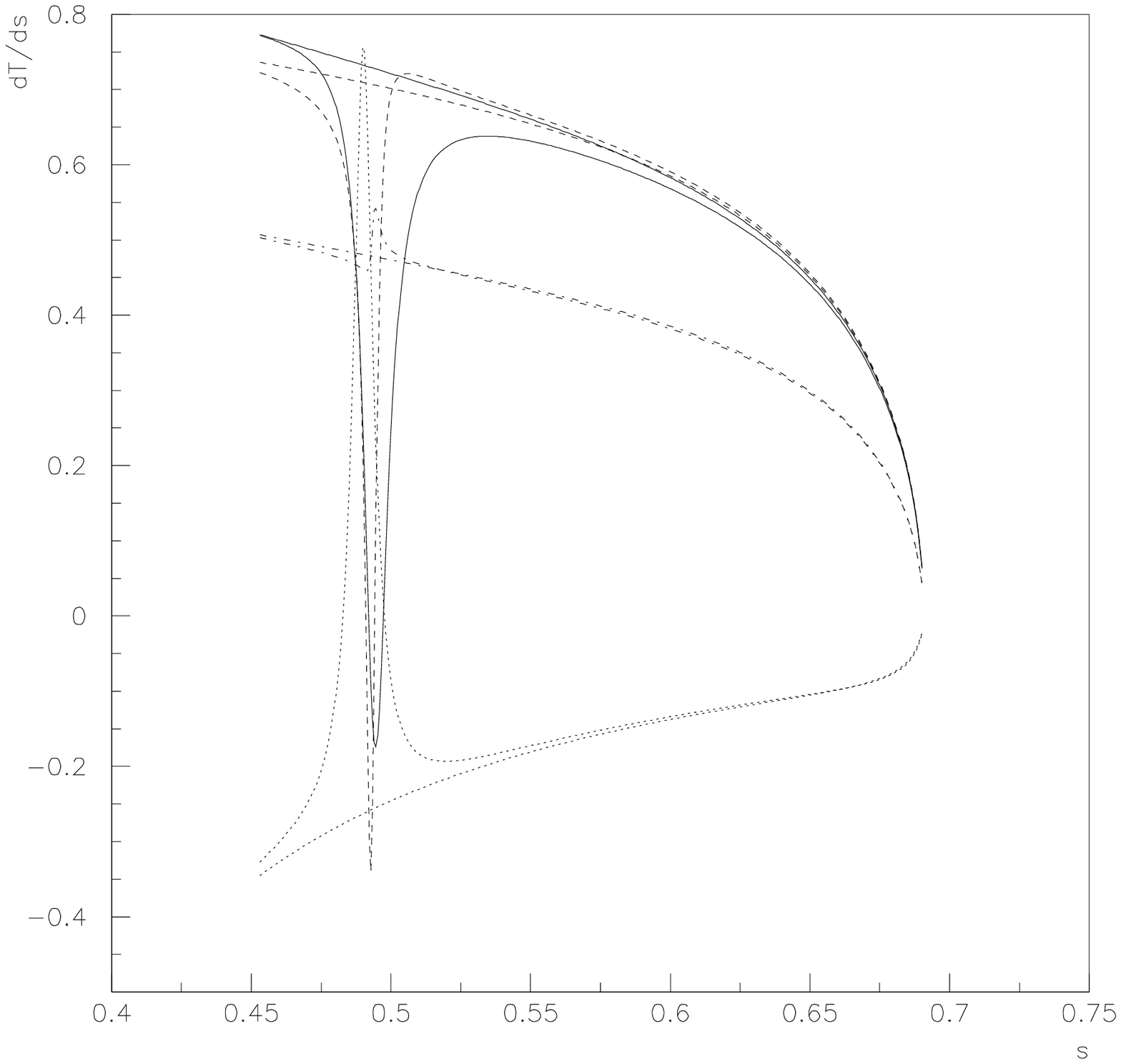,width=8.2cm}
     \mbox{ }\hfill\hspace{1cm}(d)\hfill\mbox{ }
     \end{minipage}
     \caption{\it The IMS(a), FBA(b), $P_L$(c), and $P_T$(d) of the process $\bkst$. The line conventions are the
same as given in the legend of Fig 1.}
\label{fig:kstan}
\end{figure}

In Fig. \ref{fig:kstan}, the IMS, FBA, longitudinal and transverse polarizations of the
$\bkst$ are depicted. The effect of NHBs does show in great relief.  It is worth to note that
IMS, FBA, and lepton polarizations for $\bksll$ in MSSM without including the contributions of NHBs are also significantly
diffident from those in SM, while for $\bkll$ they have little differences from those in SM. Therefore, compared to the
process $\bkll$, more precise measurements for $\bksll$ are needed  in order to single out the contributions of NHBs.

Normal polarizations for both $\bkll$ and $\bksll$ are small and can be neglected because the imaginary
parts of Wilson coefficients are small in SUSY models without CP violating phases which are implicitly
assumed in the paper.

The behavior of IMS(a), FBA(b), $P_L$(c), and $P_T$(d) shown Figs 1-4 can be understood with the
formula given in the Section 3.
With Eqs. (\ref{eq:dwbpll}), (3.10) and (3.11), we see that the contributions
of NHBs are contained in the terms of $\s_1$ and $D^{\prime}$. At the high $\sh$ regions,
it is these two terms which are important. This explained the behavior of IMS
given in (a) of Fig. 1 and Fig. 2.
The Eq. (\ref{eqn:bkas}) shows that the FBA is proportional to the mass of the lepton.
For the case $\bkm$, due to smallness of the mass $\mu$, the FBA does not vanish but is hard to
be measured. While for the case $\bkt$, the mass $\tau$ is quite large and observing FBA
is relatively easy. For SUSY II, though the numerator of
FBA is comparatively large, the large IMS suppresses the value of FBA;
for SUSY III, the numerator is relatively small, but the FBA do demonstrate the
effects of NHBs more manifestly, as shown in Fig. 2(b) due to smallness of IMS.
The Eqs. (3.63) and (3.64)  show that for the case $\ell=\mu$, the contributions of NHBs
to $P_N, P_T$ are
suppressed by the mass of $\mu$. But for the case $\ell=\tau$, the contributions
of NHBs become quite manifest both for SUSY II and SUSY III.
The term with $\rp$ in Eq. (\ref{eqn:bktp}) will change its sign
when there exists relatively not too small contributions of NHBs, the fact
deduced from Eq. (3.10), that explains why the sign of $P_T$ is changed. The difference
between the case SUSY II and SUSY III is small, the reason is just the same as stated in
the analysis of FBA.

Since the terms incorporating the contributions of NHBs is proportional to
$\la$ as shown in Eq. (\ref{eq:dwbvll}), which approaches zero at high $\sh$
regions; while at small $\sh$ regions, the effects
of NHBs are dwarfed by the other contributions. Therefore, only when
 $C_{Q_i}$ are quite large could effects of NHBs be manifest,
as shown in Fig. 3(a) and Fig. 4(a). According to the
Eq. (\ref{eqn:bksas}), at high $\sh$ regions, the effects of NHBs would be
suppressed by $\la$ and $1-\sh-\mvh^2$. The same suppression mechanism exists
for $P_L$. This suppression mechanism explains the fact that the processes $\bksll$
are not sensitive to the effects of NHBs. However, when there exist large contributions of NHBs, the
sign of $P_T$ will be changed, as indicated in both Fig. 3(d) and Fig. 4(d).

\begin{table}
\begin{center}
\begin{tabular}{|c|c|c|c|c|c|c|c|c|}
\hline
\multicolumn{2}{|c|}{{\scriptsize model}}&
\multicolumn{1}{|c|}{{\scriptsize A}} &
\multicolumn{1}{c|}{{\scriptsize B}} &
\multicolumn{1}{c|}{{\scriptsize C}} &
\multicolumn{1}{c|}{{\scriptsize D}} &
\multicolumn{1}{c|}{{\scriptsize E}} &
\multicolumn{1}{c|}{{tot(SD)}} &
\multicolumn{1}{c|}{{tot(SD+LD)}} \cr
\hline
{SM} &LCSR&$0.353$ &$54.707$ &$0.032$ &$4.566$ &$0.076$ &$0.573$ &$59.736$ \\
\cline{2-8}\cline{9-9}
&SVZ&$0.215$ &$22.918$ &$0.015$ &$1.593$ &$0.026$ &$0.299$ &$24.767$\\
\hline
{SUSY I} &LCSR&$0.425$ &$54.723$ &$0.037$ &$4.576$ &$0.086$ &$0.675$ &$59.847$ \\
\cline{2-8}\cline{9-9}
&SVZ&$0.179$ &$22.910$ &$0.011$ &$1.586$ &$0.019$ &$0.236$ &$24.704$\\
\hline
{SUSY II} &LCSR&$0.556$ &$54.865$ &$0.131$ &$4.833$ &$0.849$ &$2.067$ &$61.233$ \\
\cline{2-8}\cline{9-9}
&SVZ&$0.348$ &$23.009$ &$0.068$ &$1.726$ &$0.321$ &$1.002$ &$25.473$\\
\hline
{SUSY III} &LCSR&$0.429$ &$54.727$ &$0.040$ &$4.584$ &$0.109$ &$0.717$ &$59.889$ \\
\cline{2-8}\cline{9-9}
&SVZ&$0.181$ &$22.912$ &$0.012$ &$1.590$ &$0.028$ &$0.255$ &$24.723$\\
\hline
\end{tabular}
\caption{Partial decay widths for $\bkm$. LCSR means the approach light-cone QCD sum rules,
SVZ means the SVZ QCD sum rule \cite{cfss}.
Character A means the region ($\sh_0 , {({\hat m_\psi} -{\hat \delta} )}^2 $),
B (${({\hat m_\psi} -{\hat \delta})}^2 , {({\hat m_\psi} +{\hat \delta}
)}^2 $), C (${({\hat m_\psi} +{\hat \delta})}^2 ,  {({\hat m_{\psi'}}
-{\hat \delta} )}^2 $), D(${({\hat m_{\psi'}} -{\hat \delta})}^2 , {({\hat m_{\psi'}}
+{\hat \delta} )}^2 $) and E (${({\hat m_{\psi'}} +{\hat \delta})}^2 ,  \sh_{\rm
max}^2$). The unit is $\Gamma_B \times 10^{-6}$, which is $4.22\times 10^{-19}$ GeV.
$\delta$ is selected to be $0.2$ GeV. ${\hat \delta}$ is normalized with $M_B$}
\end{center}
\label{tab:pdw1}
\end{table}

\begin{table}
\begin{center}
\begin{tabular}{|c|c|c|c|c|c|c|c|c|}
\hline
\multicolumn{2}{|c|}{{\scriptsize model}}&
\multicolumn{1}{|c|}{{\scriptsize A}} &
\multicolumn{1}{c|}{{\scriptsize B}} &
\multicolumn{1}{c|}{{\scriptsize C}} &
\multicolumn{1}{c|}{{\scriptsize D}} &
\multicolumn{1}{c|}{{\scriptsize E}} &
\multicolumn{1}{c|}{{tot(SD)}} &
\multicolumn{1}{c|}{{tot(SD+LD)}} \cr
\hline
{SM} &LCSR&$0.930$ &$83.257$ &$0.141$ &$9.976$ &$0.258$ &$1.882$ &$94.562$ \\
\cline{2-8}\cline{9-9}
&SVZ&$2.943$ &$111.278$ &$0.147$ &$7.504$ &$0.137$ &$3.639$ &$122.008$\\
\hline
{SUSY I} &LCSR&$1.627$ &$83.402$ &$0.198$ &$10.085$ &0.330$$ &$2.915$ &$95.64$ \\
\cline{2-8}\cline{9-9}
&SVZ&$4.517$ &$111.423$ &$0.183$ &$7.552$ &$0.149$ &$5.291$ &$123.825$\\
\hline
{SUSY II} &LCSR&$1.178$ &$83.431$ &$0.234$ &$10.164$ &$0.352$ &$2.677$ &$95.360$ \\
\cline{2-8}\cline{9-9}
&SVZ&$2.801$ &$111.292$ &$0.156$ &$7.525$ &$0.145$ &$3.522$ &$121.918$\\
\hline
{SUSY III} &LCSR&$1.631$ &$83.407$ &$0.201$ &$10.092$ &$0.334$ &$2.938$ &$95.664$ \\
\cline{2-8}\cline{9-9}
&SVZ&$4.518$ &$111.425$ &$0.184$ &$7.553$ &$0.150$ &$5.296$ &$123.830$\\
\hline
\end{tabular}
\caption{Partial decay widths for $\bksm$. Other conventions can be found in Table 5.}
\end{center}
\label{tab:pdw2}
\end{table}

\begin{table}
\begin{center}
\begin{tabular}{|c|c|c|c|c|c|}
\hline
\multicolumn{1}{|c|}{{\scriptsize model}}&
\multicolumn{1}{|c|}{{\scriptsize }}&
\multicolumn{1}{|c|}{{\scriptsize A'}} &
\multicolumn{1}{c|}{{\scriptsize B'}} &
\multicolumn{1}{c|}{{tot(SD)}} &
\multicolumn{1}{c|}{{tot(SD+LD)}} \cr
\hline
{SM} &LCSR&$1.884$ &$0.094$ &$0.132$ &$1.978$ \\
\cline{2-6}
&SVZ&$0.659$ &$0.036$ &$0.054$ &$0.695$ \\
\hline
{SUSY I} &LCSR&$1.884$ &$0.086$ &$0.131$ &$1.970$ \\
\cline{2-6}
&SVZ&$0.655$ &$0.025$ &$0.038$ &$0.680$ \\
\hline
{SUSY II} &LCSR&$2.022$ &$1.496$ &$1.674$ &$3.519$ \\
\cline{2-6}
&SVZ&$0.726$ &$0.552$ &$0.637$ &$1.278$ \\
\hline
{SUSY III} &LCSR&$1.874$ &$0.094$ &$0.129$ &$1.968$ \\
\cline{2-6}
&SVZ&$0.651$ &$0.026$ &$0.035$ &$0.677$ \\
\hline
\end{tabular}
\caption{Partial decay widths of $\bkt$.
Character A' means ($\sh_0, {({\hat m_\psi} -{\hat \delta})}^2$),
B' means (${({\hat m_{\psi'}}+{\hat \delta})}^2, \sh_{\rm max}$).
The unit is $\Gamma_B \times 10^{-6}$, which is $4.22\times 10^{-19}$ GeV.}
\label{tab:pdw3}
\end{center}
\end{table}

\begin{table}
\begin{center}
\begin{tabular}{|c|c|c|c|c|c|}
\hline
\multicolumn{1}{|c|}{{\scriptsize model}}&
\multicolumn{1}{|c|}{{\scriptsize }}&
\multicolumn{1}{|c|}{{\scriptsize A'}} &
\multicolumn{1}{c|}{{\scriptsize B'}} &
\multicolumn{1}{c|}{{tot(SD)}} &
\multicolumn{1}{c|}{{tot(SD+LD)}} \cr
\hline
{SM} &LCSR&$4.045$ &$0.096$ &$0.183$ &$4.141$ \\
\cline{2-6}
&SVZ&$3.029$ &$0.048$ &$0.102$ &$3.076$ \\
\hline
{SUSY I} &LCSR&$4.088$ &$0.173$ &$0.327$ &$4.261$ \\
\cline{2-6}
&SVZ&$3.052$ &$0.072$ &$0.159$ &$3.124$ \\
\hline
{SUSY II} &LCSR&$4.148$ &$0.266$ &$0.460$ &$4.413$ \\
\cline{2-6}
&SVZ&$3.054$ &$0.084$ &$0.167$ &$3.138$ \\
\hline
{SUSY III} &LCSR&$4.078$ &$0.168$ &$0.312$ &$4.246$ \\
\cline{2-6}
&SVZ&$3.050$ &$0.071$ &$0.156$ &$3.121$ \\
\hline
\end{tabular}
\caption{Partial decay widths of $\bkst$. Other conventions can be found at
Table 7.}
\end{center}
\label{tab:pdw4}
\end{table}

The partial decay widths (PDW) are listed in Tables. 5,6,7,and 8. We see that at the high $\sh$ region, for the process
$B\rightarrow K l^+ l^-$, l=$\mu,\tau$, the
contributions of NHBs do show up, as expected. For $B\rightarrow K^{*} l^+l^-$, the effects of NHBs in the high $\hat{s}$
region is signifiacnt when l=$\tau$ while they are small for l=$\mu$.
It can be read out from these four table that
the results are consistent with the Fig. \ref{fig:kman}(a),\ref{fig:ktan}(a),
\ref{fig:ksman}(a), and \ref{fig:kstan}(a).
In order to estimate the theoretical uncertainty brought by the methods
calculating the weak form factors, we use the form factors calculated with LCSR and SVZ QCD sum
rules (SVZ) method \cite{cfss}.
For $\bkll$, PDWs calculated with form factors obtained by SVZ method
is 50\% of those by LCSR approach; while for $\bksll$, PDWs increase 100\% or so.
We see that at low $\sh$ regions the theoretical uncertainty can reach from 100\% to 200\%.
Another point worthy of mention is that the contribution of resonences domainate the integerated decay width, as had been pointed out in \cite{liu}.

\section{Conclusion}
We have calculated invariant mass spectrum, FBAs, and lepton polarizations for $\bkll$ and $\bksll$
l=$\mu,\tau$ in SUSY theories. In particular, we have analyzed the effects of NHBs to these processes. It is shown that the effects
of the NHBs  to $\bkt$ and $\bkst$ in some regions of parameter space of SUSY models
are considerable and remarkable. The reason lies in the mass of the $\tau$, which can
magnify the effects of NHBs and can be see through from the related formula.
The numerical results imply that there still exist possiblities to observe the effects of NHB in $\bkm$ and $\bksm$ through
IMS, FBA and lepton polarizations of these processes. In particular, for $B\rightarrow K \mu^+\mu^-$
in the case of SUSY II,   the partial width in the high $\hat{s}$ where short distance physics dominants  can be enhanced by a
factor of 12 compared to SM. Our analysis also show that
the theoretical uncertainties brought in calculating of weak form factors
are quite large. But the effects of NHBs will not be washed out and can stand out in some regions of the parameter space in
MSSM. If only partial widths are
measured, it is difficult to observe the effects of NHBs except for the decay $B\rightarrow K \tau^+\tau^-$. However, the conbined
analysis of IBS, FBA, and lepton polarizations can provide usefull knowledge to look for SUSY.
Finally, we would like to point out that FBA for $B\rightarrow K l^+l^-$ vanishes (or, more precisely, is neglegiblly small)
in SM and it does not vanish in 2HDM and SUSY models with large tan$\beta$ due to the contributions of NHBs. However, only
in SUSY models and for l=$\tau$ it is large enough to be observed in B factories in the near future.

\section*{Acknowledgements}
One of the Authors ( Q.S YAN) would like to thank Dr. W.J. HUO for his help during the work. This work was support in part
by the National Nature Science Fundation of China.

\end{document}